\documentclass[reqno,11pt]{amsart}
\usepackage[linktocpage]{hyperref}
\usepackage[utf8]{inputenc}
\usepackage{graphicx}
\usepackage{slashed}
\usepackage{amscd}
\usepackage{amssymb}
\usepackage{esint}
\usepackage{enumitem}
\usepackage{dsfont}
\usepackage[mathscr]{eucal}
\usepackage{upgreek}
\usepackage{epsfig}
\usepackage{pst-all}

\numberwithin{equation}{section}
\allowdisplaybreaks[1]

\title[A Notion of Entropy for Causal Fermion Systems]{A Notion of Entropy for Causal Fermion Systems}
\author[F.\ Finster]{Felix Finster \\ \\ March/August 2021}
\address{Fakult\"at f\"ur Mathematik  Universit\"at Regensburg  D-93040 Regensburg  Germany}
\email{finster@ur.de}

\newtheorem{Def}{Definition}[section]
\newtheorem{Thm}[Def]{Theorem}
\newtheorem{Prp}[Def]{Proposition}
\newtheorem{Lemma}[Def]{Lemma}
\newtheorem{Remark}[Def]{Remark}

\newcommand{\Thanks}{\vspace*{.5em} \noindent \thanks}
\newcommand{\beq}{\begin{equation}}
\newcommand{\eeq}{\end{equation}}
\newcommand{\Proof}{\begin{proof}}
	\newcommand{\QED}{\end{proof} \noindent}
\newcommand{\QEDrem}{\ \hfill $\Diamond$}

\newcommand{\la}{\langle}
\newcommand{\ra}{\rangle}

\newcommand{\C}{\mathbb{C}}
\newcommand{\R}{\mathbb{R}}
\newcommand{\1}{\mbox{\rm 1 \hspace{-1.05 em} 1}}
\newcommand{\Z}{\mathbb{Z}}
\newcommand{\N}{\mathbb{N}}

\DeclareMathOperator{\tr}{tr}
\renewcommand{\O}{{\mathscr{O}}}
\renewcommand{\L}{{\mathcal{L}}}
\newcommand{\Sact}{{\mathcal{S}}}

\newcommand{\scrt}{{\mathfrak{t}}}

\newcommand\B{{\mathscr{B}}}
\newcommand{\U}{\text{\rm{U}}}

\renewcommand{\H}{\mathscr{H}}

\newcommand{\Lin}{\text{\rm{L}}}
\newcommand{\F}{{\mathscr{F}}}

\newcommand{\D}{\mathscr{D}}

\DeclareMathOperator{\supp}{supp}

\newcommand{\scrM}{\mycal M}
\newcommand{\scrN}{\mycal N}
\newcommand{\itemD}{\item[{\raisebox{0.125em}{\tiny $\blacktriangleright$}}]}
\newcommand{\scrU}{{\mathscr{U}}}

\newcommand{\s}{\mathfrak{s}}

\newcommand{\bitem}{\begin{itemize}[leftmargin=2.5em]}
\newcommand{\eitem}{\end{itemize}}

\newcommand{\e}{{\text{\rm{e}}}}

\newcommand{\G}{\mathscr{G}}
\newcommand{\Fock}{{\mathcal{F}}}

\newcommand{\fermi}{{\mathrm{\tiny{f}}}}

\newcommand{\x}{\mathbf{x}}
\newcommand{\y}{\mathbf{y}}

\DeclareFontFamily{OT1}{rsfso}{}
\DeclareFontShape{OT1}{rsfso}{m}{n}{ <-7> rsfso5 <7-10> rsfso7 <10-> rsfso10}{}
\DeclareMathAlphabet{\mycal}{OT1}{rsfso}{m}{n}

\begin{document}
\maketitle

\begin{abstract}
A notion of entropy is introduced for causal fermion systems.
This entropy is a measure of the state of disorder of a causal fermion system
at a given time compared to the vacuum.
The definition is given both in the finite and infinite-dimensional
settings. General properties of the entropy are analyzed.
\end{abstract}

\tableofcontents

\section{Introduction} \label{secintro}
Entropy is a measure for the disorder of a physical system.
There are various notions of entropy, like the entropy in classical statistical mechanics
as introduced by Boltzmann and Gibbs, the Shannon and R{\'e}nyi entropies in information theory
or the von Neumann entropy for quantum systems.
Moreover, to a subsystem of a quantum system one can associate a corresponding entanglement entropy.
In this paper, we shall complement these concepts by a notion of entropy for causal fermion systems.
This entropy quantifies the state of disorder of a causal fermion system at any given time compared to a causal fermion system
describing the vacuum. This entropy is positive, and it vanishes in the vacuum.
Its mathematical structure formally resembles an entropy in that a logarithm is involved.
Its detailed form, however, is quite different from other entropies and connects to the
specific mathematical structures of a causal fermion system. In fact, our entropy
is closely related to the partition function as introduced in~\cite{fockfermionic} for the construction
of the quantum state of a causal fermion system.
Our entropy should be regarded as a quantum entropy. In particular, it gives rise to
a corresponding entanglement entropy.

The theory of {\em{causal fermion systems}} is a recent approach to fundamental physics
(see the basics in Section~\ref{secprelim}, the reviews~\cite{review, dice2014}, the 
textbooks~\cite{cfs, intro} or the website~\cite{cfsweblink}).
In this approach, spacetime and all objects therein are described by a measure~$\rho$
on a set~$\F$ of linear operators on a Hilbert space~$(\H, \la .|. \ra_\H)$. 
The physical equations are formulated via the so-called {\em{causal action principle}},
a nonlinear variational principle where an action~$\Sact$ is minimized under variations of the measure~$\rho$.
{\em{Spacetime}}~$M$ is defined to be the support of this measure,
\beq \label{Mdef}
M := \supp \rho \subset \F \:.
\eeq
Our general strategy is to ``compare'' two causal fermion
systems~$(\H, \F, \rho)$ and~$(\tilde{\H}, \tilde{\F}, \tilde{\rho})$ which can be viewed as
describing the vacuum and an interacting system, respectively.
We assume that both measures are minimizers of the causal action.
Before we can ``compare'' the causal fermion systems, we must
identify the Hilbert spaces~$\H$ and~$\tilde{\H}$.
The fact that this identification is not canonical gives rise to an intrinsic freedom in choosing a unitary
transformation~$\scrU$. The basic idea behind our notion of entropy is to measure the state of disorder
of the system by analyzing the fluctuations of a certain functional when this unitary transformation is varied.
The functional we consider is the so-called {\em{nonlinear surface layer integral}} as introduced and
used in~\cite{fockbosonic, fockfermionic} for the construction of quantum states, which takes the form
\beq \label{OSInonlin}
\gamma^{\tilde{\Omega}, \Omega}(\tilde{\rho}, \scrU \rho) :=
\bigg( \int_{\tilde{\Omega}} d\tilde{\rho}(x) \int_{M \setminus \Omega} d\rho(y) - 
\int_{\tilde{M} \setminus \tilde{\Omega}} d\tilde{\rho}(x) \int_{\Omega} d\rho(y) \bigg) 
\L \big( x,\scrU y \scrU^{-1} \big) \:,
\eeq
where the sets~$\tilde{\Omega} \subset \tilde{M} := \supp \tilde{\rho}$
and~$\Omega \subset M := \supp \rho$ can be thought of as the pasts of the time
for which we want to compute the entropy (for details see the preliminaries
in Section~\ref{secosinonlin}).
In order to disregard the unitary transformations which describe time evolutions, we restrict attention
to a subset of unitary transformations denoted by~$\G^{t_0}$, defined by an equation of the form
\beq \label{timefix}
\gamma^{\Omega, \Omega}(\rho, \scrU \rho) = 0 \:.
\eeq
It is one of our tasks to construct a normalized integration measure~$\mu_\G^{t_0}$ on the set~$\G^{t_0}$.
Then, given the set~$\tilde{\Omega} \subset M$,
we choose a unitary operator~$h \in \U(\H)$ and a set~$\tilde{\Omega}' \subset \tilde{M}$
satisfying the constraints
\beq \label{constraints}
\fint_{ \G^{t_0}} \gamma^{\tilde{\Omega}', \Omega} \big(\tilde{\rho}, h \scrU \rho \big)\:
d\mu^{t_0}_\G(\scrU) = 0 
= \fint_{ \G^{t_0}} \gamma^{\tilde{\Omega}, \Omega} \big(\tilde{\rho}, h \scrU \rho \big)\:
d\mu^{t_0}_\G(\scrU)
\eeq
(where $\fint$ denotes a normalized integral)
and consider the logarithm of the integral of the exponential of the nonlinear surface layer integral,
\beq \label{expint}
{\mathscr{S}} \big(h, \tilde{\Omega}' \big) = \log \fint_{ \G^{t_0}} e^{\beta \gamma^{\tilde{\Omega}', \Omega} \big(\tilde{\rho}, h \scrU \rho \big)}\:
d\mu^{t_0}_\G(\scrU)\:,
\eeq
where~$\beta$ is a given real parameter.
Then, using the first constraint in~\eqref{constraints} and applying
Jensen's inequality together with the fact that the exponential is convex, one
finds that the expression~\eqref{expint} is non-negative, and it is zero if and only if the integrand is constant for
almost all~$\scrU \in \G^{t_0}$. This leads us to define the entropy by taking the infimum over
the remaining freedom, i.e.\
\beq \label{Saux}
{\mathscr{S}}(\tilde{\Omega}) := \inf_{(h, \tilde{\Omega}')} {\mathscr{S}} \big(h, \tilde{\Omega}' \big) \:,
\eeq
where the pairs~$(h, \tilde{\Omega}')$ must satisfy the constraints~\eqref{constraints}.

Making this idea mathematically precise,
one encounters several technical difficulties which we treat in various situations.
In the {\em{finite-dimensional setting}}
\beq \label{fdef}
f:= \dim \H = \dim \tilde{\H} < \infty \:,
\eeq
the unitary group~$\G := \U(\H)$ is compact. In general, however, it is not clear
how to treat the unitary transformations describing time translations.
This difficulty can be bypassed by working with time strips and taking the limit when the time interval
of these strips tends to zero (Section~\ref{secfinitegen}). This leads to a general notion of entropy (Definition~\ref{defentropyfinite})
which can be adapted even to non-continuum or discrete spacetimes (Remark~\ref{remdiscrete}).
In order to be able to divide out the translations by an equation of the from~\eqref{timefix},
we assume that the vacuum measure~$\rho$ is {\em{static}} (Section~\ref{secfinitestatic}).
In this case, the set~$\G^{t_0}$ can be defined and, under general regularity assumptions,
it is a submanifold of~$\G$ (Lemma~\ref{lemmasub}). On the other hand, the static finite-dimensional setting
suffers from the shortcoming that the support of the measure, and consequently also the
orbits of the time translations, are compact.
In other words, the vacuum spacetime~$M \simeq S^1 \times N$ is necessarily {\em{time-periodic}}. As a consequence,
the above surface layer integrals are meaningful only if one multiplies~$\rho$ by a cutoff function~$\eta$
in time. The resulting drawback is that the new measure~$\eta \rho$ is no longer static,
and it is no longer a minimizer of the causal action principle.
This leads to complications and makes it necessary to impose additional technical conditions
(Definitions~\ref{defttr} and~\ref{defttrtilde}).

These shortcomings and technical complications disappear in the {\em{infinite-di\-men\-sio\-nal setting
with static vacuum}} (Section~\ref{secinfinite}),
in which case the orbits of the time evolution can be non-compact, so that the vacuum spacetime
has the product form~$M \simeq \R \times N$. However, the difficulty arises that also
the unitary group~$\U(\H)$ is non-compact. Thus, in order to make mathematical sense of integrals over the group,
we need to exhaust~$\U(\H)$ by compact subgroups~$\G$ and take the infimum over all exhaustions.

For clarity of presentation, we begin in the finite-dimensional setting and give all definitions there.
Denoting configurations~$(h, \tilde{\Omega}')$ which realize the minimum in~\eqref{Saux} as being
{\em{optimal}}, we prove the existence of such optimal configurations (Proposition~\ref{prpoptimal})
and work out corresponding optimality conditions (Proposition~\ref{lemmaELoptimal}).
The remaining question is whether the optimal set~$\tilde{\Omega}'$ is unique.
This is an important question which leads to an interesting interplay between the
geometry of spacetime and the action of the group of unitary transformations~$\scrU$
on the Hilbert space~$\H$. 
However, this question is too difficult to answer in full generality.
But at least, we succeed in proving under suitable assumptions that in the case~$\rho=\tilde{\rho}$
of the vacuum spacetime and choosing~$(h, \tilde{\Omega}) = (\e, \Omega)$, second variations of
the entropy around~$\Omega'=\Omega$ are strictly positive if~$|\beta|$
is chosen sufficiently large (see Theorem~\ref{thmunique}
and the preceding motivation of the assumptions).
This suggests that also the set~$\tilde{\Omega}'$ should be unique, provided that~$\tilde{\rho}$
and~$\tilde{\Omega}$ are small perturbations of~$\rho$ and~$\tilde{\Omega}$, respectively.
However, for technical simplicity we do not make mathematically precise
what ``smallness'' of the perturbation means.
We remark that the resulting set~$\tilde{\Omega}'$ can be used as the starting point for the construction of the quantum state
in~\cite{fockfermionic}. In particular, one can define the partition function~$Z$ by
\beq \label{Zdef}
Z = \int_\G  e^{\beta \gamma^{\tilde{\Omega}', \Omega} \big(\tilde{\rho}, h \scrU \rho \big)}\: d\mu_\G(\scrU) \:.
\eeq
In this sense, the constructions here
complement and complete those in~\cite{fockfermionic}.

Next, we move on to the infinite-dimensional setting and introduce a corresponding entropy
(Definition~\ref{defSinf}).
However, in order to avoid overly technical constructions, we do not analyze the optimal
configurations in the infinite-dimensional setting.

The paper is organized as follows.
In Section~\ref{secprelim} we provide the necessary preliminaries on causal fermion systems
In Section~\ref{secfinite} we consider the case that the Hilbert space is finite-dimensional.
We introduce the entropy, first in general and then in the static setting, and
study its properties. Moreover, we analyze properties of the optimal configurations~$(h, \tilde{\Omega}')$
and study existence and uniqueness.
In Section~\ref{secinfinite} we extend the concepts and definitions to the infinite-dimensional setting
with static vacuum.
In Section~\ref{secloc} we introduce the entropy of a spatial subregion and also define a corresponding
entanglement entropy. 
In Section~\ref{secoutlook} we conclude the paper by explaining the significance of the parameter~$\beta$ in~\eqref{expint} (Remark~\ref{rembeta}) and by comparing our entropy to other notions of entropy
(Remark~\ref{remcompare}).

\section{Preliminaries} \label{secprelim}
\subsection{A Few Basics on Causal Fermion Systems} \label{seccapbasics}
This section is intended for readers who are not familiar with causal fermion systems.
Our presentation has similarities to other introductions (for example~\cite[Section~2]{dice2014},
\cite[Section~1]{nrstg}, \cite[Section~1.2]{cfs} or~\cite[Section~4]{review}),
but it is streamlined towards the structures needed in the present paper.

In order to explain the basic setup, we begin with a simple example in the setting of relativistic quantum mechanics.
Let~$\scrM$ be Minkowski space and $\mu$ the standard volume measure thereon, i.e.~$d\mu = d^4x$ if $x=(x^0,x^1,x^2,x^3)$ is an inertial frame.
We consider wave functions which satisfy the Dirac equation
\begin{align}
\label{DiracEq}
\big( i \gamma ^j \partial_j + \B - m \big)\, \psi = 0 \:,
\end{align}
where~$m$ is the rest mass, $\gamma^j$ are the Dirac matrices, and~$\B$
is a potential describing an interaction.
On the Dirac solutions, we consider the usual scalar product
\begin{align}
\label{ScalProd}
( \psi | \phi )_t := \int_{t=\textrm{const}} (\overline \psi \gamma^0  \phi) (t,\vec x) \:d^3 x
\end{align}
(here~$\overline{\psi} = \psi^\dagger \gamma^0$ is the adjoint spinor, where the dagger denotes complex conjugation and transposition). If one evaluates~\eqref{ScalProd} for~$\phi=\psi$,
the integrand can be written as~$(\overline{\psi}\gamma^0\psi)(t,\vec{x}) = (\psi^\dagger \psi)(t,\vec{x})$,
having the interpretation as the probability density of the Dirac particle described by~$\psi$
to be at at the position~$\vec{x}$ at time~$t$. Due to current conservation, the integral in~\eqref{ScalProd} is time independent.

Next, we choose an ensemble of Dirac solutions~$\psi_1, \ldots, \psi_f$.
For simplicity in presentation, we restrict attention to the case~$f<\infty$ of a finite number of
Dirac wave functions, which we assume to be continuous.
It is a central idea behind causal fermion systems to describe the physical system and
to formulate its dynamical equations purely in terms of the ensemble of wave
functions~$\psi_1, \ldots, \psi_f$.
To this end, we denote the complex vector space spanned by the
wave functions~$\psi_1, \ldots, \psi_f$ by~$\H$. On~$\H$ we consider the restriction of the
scalar product~\eqref{ScalProd}, i.e.\ $\la .|. \ra_\H := ( .|. )_t|_{\H \times \H}$.
Thus~$(\H, \la .|. \ra_\H)$ is an $f$-dimensional Hilbert space,
whose vectors are represented by wave functions.
For any spacetime point~$x \in \scrM$, we now introduce the sesquilinear form
\beq \label{bxdef}
b_x :  \H \times \H \rightarrow \C \:,\qquad b_x(\psi, \phi) = -(\overline \psi \phi) (x) \:,
\eeq
which maps two solutions of the Dirac equation to their spin inner product at $x$.
The sesquilinear form $b_x$ can be represented by an operator $F(x)$ on $\H$,
which is uniquely defined by the relations
\[ \la \psi | F(x) \phi \ra_\H =b_x(\psi,\phi) \qquad \text{for all~$\psi, \phi \in \H$}\:. \]
More concretely, in an orthonormal basis~$(\psi_k)_{k = 1, \ldots,f}$ of~$\H$, the last relation can be written as
\begin{align} \label{Fdef}
\la \psi_i | F(x) \psi_j \ra_\H = - \big(\overline{\psi_i} \psi_j \big)(x) \:.
\end{align}
In physical terms, the matrix element~$-(\overline{\psi_i} \psi_j)(x)$ gives information on the correlation of the
wave functions~$\psi_i$ and~$\psi_j$ at the spacetime point~$x$.
Therefore, we refer to~$F(x)$ as the {\em{local correlation operator}} at~$x$.

Let us analyze the properties of $F(x)$. First of all, the calculation
\[ \la F(x) \,\psi \,|\, \phi \ra_\H = \overline{ \la \phi \,|\, F(x) \,\psi \,\ra_\H}
= -\overline{(\overline \phi \psi) (x)} = -(\overline \psi \phi) (x) = \la \psi \,|\, F(x) \,\phi \ra_\H \]
shows that the operator~$F(x)$ is symmetric
(where we denoted complex conjugation by a bar).
Furthermore, since the spin inner product $(\overline \psi \phi)(x)$ has signature $(2,2)$,
we know that~$b_x$ has signature $(p,q)$ with $p,q \leq 2$.
As a consequence, counting multiplicities, the operator~$F(x)$ has at most two positive and at most two negative eigenvalues. By rescaling, we arrange the the operator~$F(x)$ has trace one
(this rescaling is not of relevance for the basic understanding; it will explained after~\eqref{volconstraint} below).
It is useful to denote the set of all symmetric linear operators on~$\H$ which have trace one, rank at most
four and (counting multiplicities) have at most two positive and at most two negative eigenvalues by~$\F
\subset \Lin(\H)$. Then the local correlation operator~$F(x)$ is an element of~$\F$.

Constructing the operator $F(x) \in \F$ for every spacetime point $x \in \scrM$, we
obtain the {\em{local correlation map}}
\beq \label{lcm}
F : \scrM \rightarrow \F \:,\qquad x \mapsto F(x) \:.
\eeq
This allows us to introduce a measure $\rho$ on $\F$ as follows. For any~$\Omega \subset \F$,
one takes the pre-image $F^{-1}(\Omega) \subset \scrM$ and computes its spacetime volume,
\[ \rho(\Omega) := \mu \big( F^{-1}(\Omega) \big) \:. \]
This gives rise to the so-called {\em{push-forward measure}} denoted by~$\rho = F_\ast \mu$. The $\rho$-measurable sets are defined as
the $\sigma$-algebra of all subsets of~$\F$ whose pre-image~$F^{-1}(\Omega)$
is $\mu$-measurable.

\begin{figure}
\psscalebox{1.0 1.0} 
{
\begin{pspicture}(0,-1.511712)(10.629875,1.511712)
\definecolor{colour0}{rgb}{0.8,0.8,0.8}
\definecolor{colour1}{rgb}{0.6,0.6,0.6}
\pspolygon[linecolor=black, linewidth=0.002, fillstyle=solid,fillcolor=colour0](6.4146066,0.82162136)(6.739051,0.7238436)(6.98794,0.68384355)(7.312384,0.66162133)(7.54794,0.67939913)(7.912384,0.7593991)(8.299051,0.8705102)(8.676828,0.94162136)(9.010162,0.9549547)(9.312385,0.9371769)(9.690162,0.8571769)(10.036829,0.7371769)(10.365718,0.608288)(10.614607,0.42162135)(10.614607,-0.37837866)(6.4146066,-0.37837866)
\pspolygon[linecolor=black, linewidth=0.002, fillstyle=solid,fillcolor=colour1](6.4146066,1.2216214)(6.579051,1.1616213)(6.770162,1.1127324)(6.921273,1.0905102)(7.103495,1.0816213)(7.339051,1.0549546)(7.530162,1.0638436)(7.721273,1.0993991)(7.8857174,1.1393992)(8.10794,1.2060658)(8.299051,1.2549547)(8.512384,1.3038436)(8.694607,1.3260658)(8.890162,1.3305103)(9.081273,1.3393991)(9.379051,1.3216213)(9.659051,1.2593992)(9.9746065,1.1705103)(10.26794,1.0460658)(10.459051,0.94384354)(10.614607,0.82162136)(10.610162,0.028288014)(10.414606,0.1660658)(10.22794,0.26828802)(10.010162,0.37051025)(9.663495,0.47273245)(9.356829,0.53051025)(9.054606,0.548288)(8.814607,0.54384357)(8.58794,0.5171769)(8.387939,0.48162135)(8.22794,0.44162133)(7.90794,0.34828803)(7.6946063,0.29939914)(7.485718,0.26828802)(7.272384,0.26828802)(7.02794,0.28162134)(6.82794,0.3171769)(6.676829,0.35273245)(6.543495,0.38828802)(6.4146066,0.42162135)
\pspolygon[linecolor=black, linewidth=0.002, fillstyle=solid,fillcolor=colour0](0.014606438,0.82162136)(0.3390509,0.7238436)(0.5879398,0.68384355)(0.9123842,0.66162133)(1.1479398,0.67939913)(1.5123842,0.7593991)(1.8990508,0.8705102)(2.2768288,0.94162136)(2.610162,0.9549547)(2.9123843,0.9371769)(3.290162,0.8571769)(3.6368287,0.7371769)(3.9657176,0.608288)(4.2146063,0.42162135)(4.2146063,-0.37837866)(0.014606438,-0.37837866)
\psbezier[linecolor=black, linewidth=0.04](6.4057174,0.8260658)(7.6346064,0.45939913)(7.8634953,0.8349547)(8.636828,0.92828804)(9.410162,1.0216213)(10.165717,0.7927325)(10.614607,0.42162135)
\psbezier[linecolor=black, linewidth=0.04](0.005717549,0.8260658)(1.2346064,0.45939913)(1.4634954,0.8349547)(2.2368286,0.92828804)(3.0101619,1.0216213)(3.7657175,0.7927325)(4.2146063,0.42162135)
\rput[bl](2.0101619,0.050510235){$\Omega$}
\rput[bl](8.759051,0.0016213481){\normalsize{$\Omega$}}
\psline[linecolor=black, linewidth=0.04, arrowsize=0.09300000000000001cm 1.0,arrowlength=1.7,arrowinset=0.3]{->}(1.9434953,0.85495466)(1.8057176,1.6193991)
\rput[bl](2.0946064,1.1705103){$\nu$}
\psbezier[linecolor=black, linewidth=0.02](6.4146066,0.42384356)(7.6434956,0.057176903)(7.872384,0.43273246)(8.645718,0.52606577)(9.419051,0.61939913)(10.174606,0.39051023)(10.623495,0.019399125)
\psbezier[linecolor=black, linewidth=0.02](6.410162,1.2193991)(7.639051,0.8527325)(7.86794,1.228288)(8.6412735,1.3216213)(9.414606,1.4149547)(10.170162,1.1860658)(10.619051,0.8149547)
\rput[bl](8.499051,0.9993991){\normalsize{$y$}}
\rput[bl](7.8657174,0.49273247){\normalsize{$x$}}
\psdots[linecolor=black, dotsize=0.06](8.170162,0.65273243)
\psdots[linecolor=black, dotsize=0.06](8.796828,1.1327325)
\psline[linecolor=black, linewidth=0.02](6.1146064,1.2216214)(6.103495,0.82162136)
\rput[bl](5.736829,0.8993991){\normalsize{$\delta$}}
\rput[bl](3.6146064,0.888288){$\scrN$}
\rput[bl](1.1146064,-1.4117119){$\displaystyle \int_\scrN \cdots\, d\mu_\scrN$}
\rput[bl](5.7146063,-1.511712){$\displaystyle \int_\Omega d\rho(x) \int_{M \setminus \Omega} d\rho(y)\: \cdots\:\L(x,y)$}
\psline[linecolor=black, linewidth=0.02](6.0146065,1.2216214)(6.2146063,1.2216214)
\psline[linecolor=black, linewidth=0.02](6.0146065,0.82162136)(6.2146063,0.82162136)
\end{pspicture}
}
\caption{A surface integral and a corresponding surface layer integral.}
\label{fignoether1}
\end{figure}
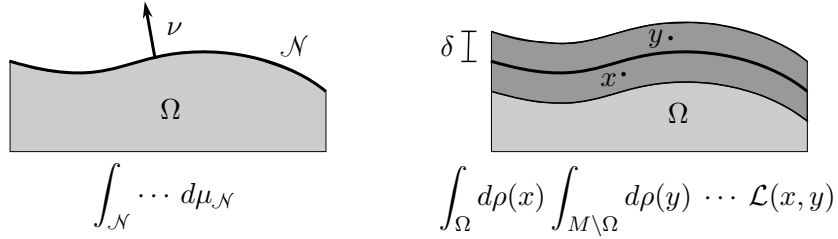

The resulting triple~$(\H, \F, \rho)$ is an example of a causal fermion system
(for the abstract definition see Definition~\ref{defcfs} below).
This example is rather special because it was obtained from Dirac wave functions
in Minkowski space. For a general causal fermion system, there is no underlying
Minkowski space, and the Dirac equation cannot be formulated.
Instead, the dynamics of a causal fermion system is described by a nonlinear variational
principle, the so-called {\em{causal action principle}}. This variational principle will be introduced
in Section~\ref{seccap} below. Here we merely make a few preliminary remarks which might help to convey the
correct physical picture behind the approach:
\bitem
\itemD The measure~$\rho$ is the basic object of the theory. It has two purposes.
First, it distinguishes which local correlations operators are ``realized'' in the physical system.
In mathematical terms, ``realized'' means that the operator lies in the support of the measure~$\rho$.
All those operators form {\em{spacetime}}~$M:= \supp \rho$.
In our above example, this connection can be understood from the local correlation map~\eqref{lcm}.
Identifying every point~$x$ of Minkowski space with its corresponding local correlation operator~$F(x)$,
spacetime becomes the image~$F(\scrM) \subset \F$ of the local correlation map.
The closure of this image indeed coincides with the support of the push-forward measure~$\rho$.
The second purpose of the measure~$\rho$ is to give a volume measure on spacetime.
Thus to a subset~$\Omega \subset M$ of spacetime we can associated its volume~$\rho(\Omega)$
(in the above example, this gives back the usual four-dimensional spacetime volume).
\itemD It is a general concept that all spacetime structures and all objects in spacetime
must be constructed from the operators in~$M$.
A particular structure encoded in these operators is the {\em{causal structure}}.
This structure is intimately related to the mathematical
form of the causal action principle, being the motivation for the name {\em{causal}}
fermion system and {\em{causal}} action principle.
\eitem
The objective of this paper is to introduce a notion of entropy for causal fermion systems.
We are facing the general issue that the entropy is a quantity to be defined in space at a fixed time.
In the setting of causal fermion systems, such spatial objects are introduced via so-called
{\em{surface layer integrals}}. We now briefly explain the basic concept.
Surface layer integrals were introduced in~\cite{noether}
as a generalization of integrals over hypersurfaces to the setting of causal fermion systems.
In general terms, a surface layer integral is a double integral of the form
\beq \label{intdouble}
\int_\Omega \bigg( \int_{M \setminus \Omega} (\cdots)\: \L(x,y)\: d\rho(y) \bigg)\, d\rho(x) \:,
\eeq
where one variable is integrated over a subset~$\Omega \subset M$, and the other
variable is integrated over the complement of~$\Omega$
(and~$(\cdots)$ stands for an unspecified differential operator acting on the Lagrangian).
In order to explain the basic idea, let us assume that the Lagrangian is of {\em{short range}} in the
sense that ~$\L$ vanishes on distances larger than~$\delta$, i.e.
\beq \label{shortrange}
d(x,y) > \delta \quad \Longrightarrow \quad \L(x,y) = 0 \:,
\eeq
where~$d \in C^0(M \times M, \R^+_0)$ is a suitably chosen distance function on~$M$.
Then the surface layer integral~\eqref{intdouble} only involves pairs~$(x,y)$ of distance at most~$\delta$,
where~$x$ is in~$\Omega$ and~$y$ is in the complement~$M \setminus \Omega$.
Thus the integral only involves points in a layer around the boundary of~$\Omega$
of width~$\delta$, i.e.
\[ x, y \in B_\delta \big(\partial \Omega \big) \:. \]
Therefore, a double integral of the form~\eqref{intdouble} can be regarded as an approximation
of a surface integral on the length scale~$\delta$, as shown in Figure~\ref{fignoether1}.
In the setting of causal variational principles, such surface layer integrals take the role of surface integrals
in Lorentzian geometry. In applications in Minkowski space or on a Lorentzian manifold,
the Lagrangian typically decays on the Compton scale $1/m$ (where~$m$ denotes again the
mass of the Dirac particles).

The differential operator $(\cdots)$ in~\eqref{intdouble} can be regarded as describing
first or second variations of the measure~$\rho$. The resulting surface layer integrals
give rise to conserved currents, the symplectic form and scalar products.
In the present paper, we need a {\em{nonlinear}} variant of surface layer integrals, where
in~\eqref{intdouble} we replace one of the measures by the measure~$\tilde{\rho}$ describing
an interacting spacetime. Here ``nonlinear'' refers to the fact that we do not assume that~$\tilde{\rho}$
is obtained from~$\rho$ by a linear or quadratic perturbation. Instead, it can be a finite, fully nonlinear
perturbation of~$\rho$. This nonlinear surface layer integral was introduced in~\cite{fockbosonic},
as will be outlined in more detail in Section~\ref{secosinonlin}.

\subsection{The Reduced Causal Action Principle} \label{seccap}
We here introduce causal fermion systems and the causal action principle
in the formulation which is most convenient for our purposes.
This formulation is obtained from the general setting as introduced in~\cite[\S1.1.1]{cfs}
by incorporating the trace and boundedness constraints into the causal action,
as will be explained in more detail at the end of this section.

\begin{Def} \label{defcfs} {\em{ 
Given a separable complex Hilbert space~$\H$ with scalar product~$\la .|. \ra_\H$
and a parameter~$n \in \N$ (the {\em{``spin dimension''}}), we let~$\F \subset \Lin(\H)$ be the set of all
symmetric operators~$A$ on~$\H$ of finite rank which have trace one,
\beq \label{fixedtrace}
\tr A = 1 \:,
\eeq
and which (counting multiplicities) have
at most~$n$ positive and at most~$n$ negative eigenvalues. On~$\F$ we are given
a positive measure~$\rho$ (defined on a $\sigma$-algebra of subsets of~$\F$).
We refer to~$(\H, \F, \rho)$ as a {\em{causal fermion system with fixed local trace}}.
}}
\end{Def} \noindent
In order to single out the physically admissible
causal fermion systems, one must formulate physical equations. To this end, we impose that
the measure~$\rho$ should be a minimizer of the causal action principle,
which we now introduce. For any~$x, y \in \F$, the product~$x y$ is an operator of rank at most~$2n$. 
However, in general it is no longer symmetric because~$(xy)^* = yx$,
and this is different from~$xy$ unless~$x$ and~$y$ commute.
As a consequence, the eigenvalues of the operator~$xy$ are in general complex.
We denote these eigenvalues counting algebraic multiplicities
by~$\lambda^{xy}_1, \ldots, \lambda^{xy}_{2n} \in \C$
(more specifically,
denoting the rank of~$xy$ by~$k \leq 2n$, we choose~$\lambda^{xy}_1, \ldots, \lambda^{xy}_{k}$ as all
the non-zero eigenvalues and set~$\lambda^{xy}_{k+1}, \ldots, \lambda^{xy}_{2n}=0$).
Given a parameter~$\kappa>0$ (which will be kept fixed throughout this paper),
we introduce the $\kappa$-Lagrangian and the causal action by
\begin{align}
\text{$\kappa$-\em{Lagrangian:}} && \L(x,y) &= \frac{1}{4n} \sum_{i,j=1}^{2n} \Big( \big|\lambda^{xy}_i \big|
- \big|\lambda^{xy}_j \big| \Big)^2 + \kappa\: \bigg( \sum_{j=1}^{2n} \big|\lambda^{xy}_j \big| \bigg)^2
\label{Lagrange} \\
\text{\em{causal action:}} && \Sact(\rho) &= \iint_{\F \times \F} \L(x,y)\: d\rho(x)\, d\rho(y) \:. \label{Sdef}
\end{align}
The {\em{reduced causal action principle}} is to minimize~$\Sact$ by varying the measure~$\rho$
under the
\beq
\text{\em{volume constraint}} \qquad \rho(\F) = \text{const} \label{volconstraint} \:.
\eeq

This variational principle is obtained from the general causal action principle as introduced in~\cite[\S1.1.1]{cfs}
as follows. Using that minimizing measures are supported on operators of constant trace
(see~\cite[Proposition~1.4.1]{cfs}), we may fix the trace of the operators. Moreover, 
by rescaling the operators according to~$x \rightarrow \lambda x$ with~$\lambda \in \R$,
one can assume without loss of generality that this trace is equal to one~\eqref{fixedtrace}.
Next, the $\kappa$-Lagrangian arises when treating the so-called boundedness constraint
with a Lagrange multiplier term. Here we slightly simplified the setting
by combining this Lagrange multiplier term with the Lagrangian right from the beginning.

This variational principle is mathematically well-posed if~$\H$ is finite-dimensional.
For the existence theory and the analysis of general properties of minimizing measures
we refer to~\cite{discrete, continuum, lagrange}.
In the existence theory one varies in the class of regular Borel measures
(with respect to the topology on~$\Lin(\H)$ induced by the operator norm),
and the minimizing measure is again in this class. With this in mind, here we always assume that
\[ 
\text{$\rho$ is a regular Borel measure}\:. \]
Given a minimizing measure~$\rho$, {\em{spacetime}}~$M$ is defined as the support of the
measure~\eqref{Mdef}, which in turn is defined as the complement of the largest open set of measure zero, i.e.
\beq \label{suppdef}
M = \supp \rho := \F \setminus \bigcup \big\{ \text{$\Omega \subset \F$ \,\big|\,
$\Omega$ is open and $\rho(\Omega)=0$} \big\} \:.
\eeq
Note that~$M$ is by definition a closed subset of~$\F$.
It is a topological space (again with the topology induced by the operator norm).

In what follows, we shall not need the specific form of the $\kappa$-Lagrangian~\eqref{Lagrange}. Instead,
we only use make use of the following of its properties:
\bitem
\item[\rm{(i)}] $\L$ is continuous.
\item[\rm{(ii)}] $\L$ is strictly positive on the diagonal,
\beq \label{strictpositive}
\L(x,x) > 0 \qquad \text{for all~$x \in \F$} \:.
\eeq
\eitem
In fact, in~\cite[Theorem~5.1]{banach} it is proven that~$\L$ is even locally H\"older continuous.
The strict positivity~\eqref{strictpositive} is quantified in~\cite[Proposition~4.3]{discrete}.

\subsection{The Euler-Lagrange Equations} \label{secEL}
A minimizer of a causal variational principle
satisfies the following {\em{Euler-Lagrange (EL) equations}}:
For a suitable value of the parameter~$\s>0$,
the lower semi-continuous function~$\ell$ defined by
\beq \label{elldef}
\ell \::\: \F \rightarrow \R\:,\qquad \ell(x) := \int_M \L(x,y)\: d\rho(y) - \s
\eeq
is minimal and vanishes in spacetime,
\beq \label{EL}
\ell|_M \equiv \inf_\F \ell = 0 \:.
\eeq
The parameter~$\s$ can be understood as the Lagrange parameter
corresponding to the volume constraint. By rescaling the measure,
one can give~$\s$ an arbitrary non-zero value. With this in mind, we keep the
parameter~$\s$ fixed throughout the paper.
For the derivation and further details on the EL equations we refer to~\cite[Section~2]{jet}.
 
\subsection{The Nonlinear Surface Layer Integral} \label{secosinonlin}
The nonlinear surface layer integral gives a way to compare two causal fermion systems
at a given time. It was first introduced in~\cite{fockbosonic} in the context of causal variational principles,
and it was used in~\cite{fockfermionic} for the construction of the quantum state of a causal fermion system.
We here recall the construction in~\cite[Section~3]{fockfermionic}.
Our starting point are two causal fermion systems~$(\H, \F, \rho)$ and~$(\tilde{\H}, \tilde{\F}, \tilde{\rho})$
which can be thought of as describing the vacuum and the interacting system, respectively.
We assume that both measures are minimizers of the causal action, and we denote the corresponding spacetimes by
\[ 
M := \supp \rho \subset \F \qquad \text{and} \qquad \tilde{M} := \supp \tilde{\rho} \subset \tilde{\F} \:. \]
Before we can get a connection between these spacetimes we must identify the
Hilbert spaces~$\H$ and~$\tilde{\H}$ by a unitary transformation denoted by~$V$,
\beq \label{Videntify}
V : \H \rightarrow \tilde{\H} \qquad \text{unitary} \:.
\eeq
Then operators in~$\tilde{\F}$ can be identified with operators in~$\F$ by the unitary transformation,
\[ 
\F = V^{-1}\, \tilde{\F}\, V \:. \]
For ease in notation, in what follows we always identify~$\H$ and~$\tilde{\H}$ via~$V$,
making it possible to always work in the Hilbert space~$\H$.
Then, given measurable subsets~$\Omega \subset M$ and~$\tilde{\Omega} \subset
\tilde{M}$, the nonlinear surface layer integral is defined by
\[ \gamma^{\tilde{\Omega}, \Omega}(\tilde{\rho}, \rho) :=
\bigg( \int_{\tilde{\Omega}} d\tilde{\rho}(x) \int_{M \setminus \Omega} d\rho(y) - 
\int_{\tilde{M} \setminus \tilde{\Omega}} d\tilde{\rho}(x) \int_{\Omega} d\rho(y) \bigg) 
\L(x,y) \:, \]
provided that the involved integrals are all finite (for a more general notion of
convergence see Definition~\ref{defic} and the subsequent analysis).

An important point to keep in mind is that the above identification~\eqref{Videntify}
of the Hilbert spaces is not canonical, but it leaves the freedom
to transform the operator~$V$ according to
\beq \label{Vtrans}
V \rightarrow V \scrU \qquad \text{with} \qquad \scrU \in \Lin(\H) \text{ unitary} \:.
\eeq
Working exclusively in the Hilbert space~$\H$, this non-uniqueness
becomes apparent in the freedom to perform unitary transformation of the vacuum measure
\[ \rho \rightarrow \scrU \rho \:, \]
where~$\scrU \rho$ is defined by
\[ 
(\scrU \rho)(\Omega) := \rho \big( \scrU^{-1} \,\Omega\, \scrU \big) 
\qquad \text{for} \qquad \Omega \subset \F \:. \]
We denote the nonlinear surface layer integral involving the unitary transformation~$\scrU$ by
\[ \gamma^{\tilde{\Omega}, \Omega}(\tilde{\rho}, \scrU \rho) :=
\bigg( \int_{\tilde{\Omega}} d\tilde{\rho}(x) \int_{M \setminus \Omega} d\rho(y) - 
\int_{\tilde{M} \setminus \tilde{\Omega}} d\tilde{\rho}(x) \int_{\Omega} d\rho(y) \bigg) 
\L \big( x, \scrU y \scrU^{-1} \big) \:. \]

\section{The Finite-Dimensional Setting} \label{secfinite}
We begin with the simplest case that the Hilbert spaces~$\H$ and~$\tilde{\H}$ are finite-dimensional.
This case has the advantage that the unitary group~$\U(\H)$ is compact.
It is a shortcoming of the finite-dimensional setting that also
spacetime is compact, thus only allowing for the description of spacetimes which are spatially compact
and have finite lifetime. With this in mind, the finite-dimensional setting is too simple for most physical
applications. Nevertheless, it is good to begin in this setting, because the definition of the entropy
is simpler and serves as a good preparation for the infinite-dimensional case.

We let~$(\H, \F, \rho)$ and~$(\tilde{\H}, \tilde{\F}, \tilde{\rho})$ be two causal fermion
systems describing the vacuum and an interacting system, respectively.
We assume that both measures are minimizers of the causal action.
Moreover, in order to allow for an identification of the Hilbert spaces~$\H$ and~$\tilde{\H}$,
we need to assume that they have the same dimension, i.e.\
\[ \dim \H = \dim \tilde{\H} =: f < \infty \:. \]
In order for the causal action principle to be well-defined, the total volume of the spacetimes must be finite.
A central object in the subsequent analysis is the group of unitary transformations of~$\H$ denoted by
\[ \G := \U(\H) \:. \]
Clearly, it is a compact Lie group. We denote the normalized Haar measure on~$\G$ by~$\mu_\G$
(for basics on compact Lie groups see for example~\cite{broecker+tomdieck}).
Next, for technical simplicity we assume that the supports of the measures are compact,
\[  M \subset \F,\; \tilde{M} \subset \tilde{\F} \quad \text{compact} \:. \]
We note that this assumption was justified in~\cite[Section~3.6]{lagrange}, where it is shown
that for minimizers of the causal action principle with~$\kappa>0$ (where~$\kappa$ is the
Lagrange parameter of the boundedness constraint), the support is indeed
bounded and thus compact.

\subsection{General Definition of the Entropy} \label{secfinitegen}
We next specify our assumptions on the causal fermion system~$(\F, \H, \rho)$ describing
the vacuum. We assume that in~$M$ there is a distinguished time function.
Since~$M$ is compact, time also takes values in a compact set.
There are two possible cases:
\bitem
\item[{\rm{(i)}}] Time takes values in a compact interval. In this case, $M$ is the topological product
\[ M = [0, T_{\max}] \times N \:. \]
Here the endpoints~$t=0$ and~$t=T_{\max}$ can be thought of as singularities of spacetimes
(the ``big bang'' and ``big crunch''). 
\item[{\rm{(ii)}}] Time takes values in~$S^1$ (a {\em{time-periodic universe}}),
leading to the topological product
\[ M=S^1 \times N \:. \]
\eitem
All the subsequent constructions apply in the same way to a spacetime of finite lifetime
and to a time-periodic universe. In order to treat both cases at once, we realize~$S^1$
as the interval~$[0,T_{\max}]$ with the endpoints identified.
Next, we assume that the measure~$\rho$ is continuous in time, meaning that it can be decomposed as
\beq \label{rhodecomp}
d\rho = dt\: d\mu_t \:,
\eeq
where the~$(\mu_t)_{t \in [0,T_{\max}]}$ are non-zero
Borel measures on~$N$. Clearly, this assumption is not satisfied for discrete
measures; see however Remark~\ref{remdiscrete} below.

We choose an intermediate time~$t \in (0,T_{\max})$ and denote its past and future by
\[ \Omega^t = [0,t] \times N \qquad \text{and} \qquad M \setminus \Omega^t = (t ,T_{\max}] \times N \:. \]
When considering surface layer integrals for this choice of~$\Omega^t$, there is the problem
that, at least in the time-periodic case, we may get contributions near the boundaries at time~$t=0$ or~$t=T_{\max}$.
In order to avoid this problem, we choose a cutoff function~$\eta \in C^\infty((0,T_{\max}), [0,1])$ which 
vanishes in a neighborhood of~$t=0$ and~$t=T_{\max}$ and is
identically equal to one in a $\delta$-neighborhood of some time~$t=t_0$, i.e.
\[ \eta|_{(t_0-\delta, t_0+\delta)} \equiv 1 \:. \]
We insert this cutoff function into the surface layer integral by setting
\beq
\begin{split}
&\gamma^{t,t'} \big( \eta \rho, \scrU (\eta \rho) \big) \\
&:= \bigg( \int_{\Omega^t} \!\!\! d(\eta \rho)(x) \int_{M \setminus \Omega^{t'}} \!\!\!\!\! d(\eta \rho)(y) - 
\int_{M \setminus \Omega^t} \!\!\!\! d(\eta \rho)(x) \int_{\Omega^{t'}} \!\!\! d(\eta \rho)(y) \bigg)
\L \big(x, \scrU y \scrU^{-1} \big) \:, \label{gammaeta}
\end{split}
\eeq
where for notational convenience we replaced the upper index~$\Omega^t$ by~$t$ and~$\Omega^{t'}$ by~$t'$.
Before going on, we point out that this setting suffers from the shortcoming that
all our results may depend on the choice of the cutoff function~$\eta$.
But this shortcoming disappears in the infinite-dimensional setting in Section~\ref{secinfinite}.

We now give the general definition of the entropy.
The unitary group~$\G$ might contain elements which describe a time evolution
(in particular, this is the case in the static setting to be considered in Section~\ref{secfinitestatic}).
These group elements should not be taken into account when integrating over~$\G$.
To this end, given sufficiently small~$\Delta t \in (0, \delta)$, we choose the set of unitary transformations
\beq \label{GtDt}
\begin{split}
\G^{t_0}(\Delta t) := \Big\{ \scrU \in \U(\H) \:\Big|&\: \exists\: t \in (t_0-\Delta t, t_0+\Delta t) \quad \text{with} \\
&\gamma^{t_0,t}\big(\eta \rho, \scrU (\eta \rho) \big)=0 \quad \text{or} \quad
\gamma^{t_0,t}\big(\eta \rho, \scrU^{-1} (\eta \rho) \big)=0
\Big\}  \:.
\end{split}
\eeq
By definition, this set is point-symmetric in the sense
\beq \label{UsymmDeltat}
\scrU \in \G^{t_0}(\Delta t) \quad \Longrightarrow \quad \scrU^{-1} \in \G^{t_0}(\Delta t) \:.
\eeq
Moreover, this set does not have measure zero:
\begin{Lemma} For any~$\Delta t>0$,
\[ \mu_\G(\G^{t_0}(\Delta t))>0 \:. \]
\end{Lemma}
\Proof Let~$\Delta t>0$. A symmetry argument in~\eqref{gammaeta} shows immediately that
\[ \gamma^{t_0, t_0} \big(\eta \rho, \eta \rho \big) = 0 \:. \]
Moreover, differentiating~\eqref{gammaeta} with respect to~$t'$ and using~\eqref{rhodecomp},
a straightforward computation yields
\beq \label{gdiff}
\frac{\partial}{\partial t'} \gamma^{t_0, t'}(\eta \rho, \eta \rho) \Big|_{t'=t_0}
= -\int_N d(\eta \mu_t)(x) \int_M \L(x, y) \:d(\eta \rho)(y) \:.
\eeq
Since~$\L$ is continuous and strictly positive on the diagonal~\eqref{strictpositive},
\beq \label{lpos}
\int_M \L(x, y) \:d(\eta \rho)(y) >0 \qquad \text{for every~$x \in M$} \:.
\eeq
This can be seen in detail as follows: Using that the Lagrangian is continuous (see
the sentence after~\eqref{strictpositive}), there is~$\delta>0$
and an open neighborhood~$U$ of~$x$ in~$\F$ such that~$\L(x,y) \geq \delta$ for all~$y \in U$.
Moreover, the set~$U$ has strictly positive measure by definition of the support~\eqref{suppdef}.
Therefore,
\[ \int_M \L(x, y) \:d(\eta \rho)(y) \geq \int_U \L(x, y) \:d(\eta \rho)(y) \geq \delta\, \rho(U) > 0 \:. \]

As a consequence of~\eqref{lpos}, also the $\mu_t$-integral of this function is strictly positive. We conclude that
\[ \frac{\partial}{\partial t'} \gamma^{t_0, t'}(\rho, \rho) \big|_{t'=t_0} < 0 \:. \]
Therefore, we can choose~$t_\pm \in (t_0-\Delta t, t_0+\Delta t)$ with
\[ t_+ < t_0 < t_- \qquad \text{and} \qquad \gamma^{t_0, t_-}(\rho, \rho) < 0 < \gamma^{t_0, t_+}(\rho, \rho) \:. \]
Using that the Lagrangian is continuous and that~$M$ is compact, the
surface layer integral~\eqref{gammaeta} is continuous in~$\scrU$. Therefore, there is
an open neighborhood~$W \subset \G$ of the neutral element $\e \in \G$ (i.e.\ the
identity operator) such that
\[ \gamma^{t_0, t_-}(\rho, \scrU \rho) < 0 < \gamma^{t_0, t_+}(\rho, \scrU \rho) \qquad \text{for all~$\scrU \in W$}\:. \]
Since the surface layer integral~\eqref{gammaeta} is also continuous in~$t'$, we can apply the intermediate
value theorem to conclude that for every~$\scrU \in W$ there is~$t \in [t_+, t_-]$ such
that~$\gamma^{t_0, t}(\rho, \scrU \rho)=0$. Using the definition of~$\G^{t_0}(\Delta t)$ in~\eqref{GtDt},
this means that~$W \subset \G^{t_0}(\Delta t)$. Hence
\[ \mu_\G \big( \G^{t_0}(\Delta t) \big) \geq \G^{t_0}(W) > 0 \:, \]
where in the last step we used that every open subset of~$\G$ has positive Haar measure.
This concludes the proof.
\QED
In view of this lemma, we may define the normalized integral over~$\G^{t_0}(\Delta t)$ by
\[ \fint_{\G^{t_0}(\Delta t)} (\cdots) \;d\mu_\G(\scrU) := \frac{1}{\mu_\G(\G^{t_0}(\Delta t))}
\int_{\G^{t_0}(\Delta t)} (\cdots) \;d\mu_\G(\scrU) \:. \]

In what follows, we let~$\tilde{\Omega}$ be a Borel subset of the interacting spacetime.
In analogy to the sets~$\Omega^t$ in the vacuum, also the set~$\tilde{\Omega}$ should be thought
of as being the ``past of a hypersurface.'' This concept will be made more precise
in Section~\ref{secexoptimal}. Here we do not need to be specific and denote the sets under consideration by
\beq \label{Pfinite}
{\mathfrak{P}}(\tilde{M}) \subset {\mathfrak{B}}(\tilde{M}) \:,
\eeq
where~$\mathfrak{B}(\tilde{M})$ are the Borel subsets of~$\tilde{M}$.
We refer to the sets in~${\mathfrak{P}}(\tilde{M})$ as {\em{past sets}}.

\begin{Def} \label{defadmissible}
The pair $(\tilde{\Omega}, h)$ with a past set~$\tilde{\Omega} \in {\mathfrak{P}}(\tilde{M})$
and~$h \in \G$ is called {\bf{admissible}} if
\beq \label{admissible}
\fint_{\G^{t_0}(\Delta t)} \gamma^{\tilde{\Omega}, t_0} \big(\tilde{\rho}, h \scrU (\eta \rho) \big)\:
d\mu_\G(\scrU) = 0 \:.
\eeq
The set of admissible pairs is denoted by
\[ {\mathscr{A}}_{\tilde{\rho}, \eta \rho}(\Delta t) \subset {\mathfrak{P}}(\tilde{M}) \times \G \:. \]
\end{Def} \noindent

\begin{Def} \label{defentropyfinite} Given a real parameter~$\beta$ and a past set~$\tilde{\Omega} \in {\mathfrak{P}}(\tilde{M})$, we define the {\bf{entropy}}~${\mathscr{S}}_{\tilde{\rho}, \eta \rho} \big(\tilde{\Omega} \big)$ by
\[ {\mathscr{S}}_{\tilde{\rho}, \eta \rho} \big(\tilde{\Omega} \big) = \liminf_{\Delta t \searrow 0}
{\mathscr{S}}_{\tilde{\rho}, \eta \rho} \big( \Delta t, \tilde{\Omega} \big) \:, \]
where
\begin{align}
{\mathscr{S}}_{\tilde{\rho}, \eta \rho} \big(\Delta t, \tilde{\Omega} \big) &:= \inf_{h \in \G \,|\, (\tilde{\Omega}, h) \in
{\mathscr{A}}_{\tilde{\rho}, \eta \rho}(\Delta t)}
{\mathscr{S}}_{\tilde{\rho}, \eta \rho} \big( \Delta t, h \big) \:, \label{SDO} \\
{\mathscr{S}}_{\tilde{\rho}, \eta \rho} \big( \Delta t, h \big) &:= \inf_{\tilde{\Omega}' \in \mathfrak{P}(\tilde{M}) \,|\, (\tilde{\Omega}', h) \in
{\mathscr{A}}_{\tilde{\rho}, \eta \rho}(\Delta t)}
\log \fint_{ \G^{t_0}(\Delta t)} e^{\beta \gamma^{\tilde{\Omega}', t_0} \big(\tilde{\rho}, h \scrU (\eta \rho) \big)}\:
d\mu_\G(\scrU) \:. \label{SDh}
\end{align}
\end{Def}

\begin{Thm} The entropy is non-negative, i.e.\ for all past sets~$\tilde{\Omega} \in {\mathfrak{P}}(\tilde{M})$,
\[ {\mathscr{S}}_{\tilde{\rho}, \eta \rho}(\tilde{\Omega}) \geq 0 \:. \]
Moreover, the entropy vanishes in the vacuum at time~$t_0$, i.e.
\beq \label{Svac}
{\mathscr{S}}_{\eta \rho, \eta \rho}(\Omega^{t_0}) = 0 \:.
\eeq
\end{Thm}
\Proof Since the exponential function is convex, Jensen's inequality (see for example~\cite[Theorem~3.3]{rudin}) yields
\[ \fint_{ \G^{t_0}(\Delta t)} e^{\beta \gamma^{\tilde{\Omega}', t_0} \big(\tilde{\rho}, h \scrU (\eta \rho) \big)}\:
d\mu_\G(\scrU) \geq \exp \bigg( \beta \fint_{ \G^{t_0}(\Delta t)} \gamma^{\tilde{\Omega}', t_0} \big(\tilde{\rho}, h \scrU (\eta \rho)
\big)\: d\mu_\G(\scrU) \bigg) = 0 \:, \]
where in the last step we used~\eqref{admissible}. We conclude that the logarithm in~\eqref{SDh}
is non-negative for any admissible pair~$(\tilde{\Omega}', h)$.
Taking the infimum over all admissible pairs and taking the limes inferior~$\Delta t \searrow 0$,
we conclude that also the entropy~${\mathscr{S}}_{\tilde{\rho}, \eta \rho} \big(\tilde{\Omega} \big)$ is non-negative.

In order to prove~\eqref{Svac}, we first note that, using the point symmetry~\eqref{UsymmDeltat}
together with the unitary invariance of the Lagrangian and the anti-symmetry of the nonlinear surface layer integral,
\begin{align*}
&\fint_{\G^{t_0}(\Delta t)} \gamma^{t_0, t_0} \big(\eta \rho, \scrU (\eta \rho) \big)\: d\mu_\G(\scrU) = \fint_{\G^{t_0}(\Delta t)} \gamma^{t_0, t_0} \big(\scrU^{-1}(\eta \rho), \eta \rho \big)\: d\mu_\G(\scrU) \\
&= -\fint_{\G^{t_0}(\Delta t)} \gamma^{t_0, t_0} \big(\eta \rho, \scrU^{-1}(\eta \rho) \big)\: d\mu_\G(\scrU)
= -\fint_{\G^{t_0}(\Delta t)} \gamma^{t_0, t_0} \big(\eta \rho, \scrU(\eta \rho) \big)\: d\mu_\G(\scrU) \:.
\end{align*}
We conclude that, choosing~$\tilde{\rho}=\eta \rho$, the pair~$(\Omega^{t_0}, \e)$ is admissible.
As a consequence, by definition of the infimum, we know that for any~$\Delta t \leq \delta$,
\beq \label{Ses}
{\mathscr{S}}_{\eta \rho, \eta \rho}(\Omega^{t_0}) \leq \log \fint_{ \G^{t_0}(\Delta t)}
e^{\beta \gamma^{t_0, t_0} \big(\eta \rho, \scrU (\eta \rho) \big)}\: d\mu_\G(\scrU) \:.
\eeq
Hence our remaining task is to estimate the surface layer
integral~$|\gamma^{t_0, t_0}(\rho, \scrU (\eta \rho))|$ from above.
According to~\eqref{GtDt}, there is~$t \in (t_0-\Delta t, t_0+\Delta t)$
with~$\gamma^{t_0,t}(\eta \rho, \scrU (\eta \rho))=0$. Thus,
using the mean value inequality,
\begin{align}
\big| \gamma^{t_0, t_0} \big(\eta \rho, \scrU (\eta \rho) \big) \big|
&= \big| \gamma^{t_0, t_0} \big(\eta \rho, \scrU (\eta \rho) \big) - \gamma^{t_0, t} \big(\eta \rho, \scrU (\eta \rho) \big) \big|
\notag \\
&\leq \Delta t\: \sup_{t \in (t_0-\Delta t, t_0+\Delta t)} \bigg|
\frac{\partial}{\partial t} \gamma^{t_0, t} \big(\eta \rho, \scrU (\eta \rho) \big) \bigg| \:. \label{gttes}
\end{align}
Using~\eqref{gdiff} together with the compactness of~$M$ and~$\G$ and the continuity of~$\L$,
one sees that the last time derivative is uniformly bounded, i.e.\ there is~$C>0$ such that
\[ \Big| \frac{\partial}{\partial t} \gamma^{t_0, t} \big(\eta \rho, \scrU (\eta \rho) \big) \Big|
\leq C \qquad \text{for all~$t \in (t_0-\delta, t_0+\delta)$ and~$\scrU \in \G$}\:. \]
Employing this inequality in~\eqref{gttes} and~\eqref{Ses}, we infer that
\[ {\mathscr{S}}_{\eta \rho, \eta \rho}(\Omega^{t_0}) \leq \log \exp \big( C\, \Delta t \big)
= C\, \Delta t \:. \]
Since~$\Delta t$ can be chosen arbitrarily small, we obtain the result.\QED

\begin{Remark} \label{remdiscrete} {\bf{(entropy for non-continuum spacetimes)}} {\em{
A discrete spacetime is described by a measure~$\rho$ with discrete support
(like for example a weighted counting measure as considered in~\cite{support}).
More generally, if spacetime has a non-continuum or non-regular structure,
the measure cannot be decomposed in the form~\eqref{rhodecomp}, so that
the above definition of entropy in Definition~\ref{defentropyfinite} cannot be used.
The basic problem is that, in a non-continuum spacetime, it is not sensible to 
work with a continuous time parameter and to take the limit~$\Delta t \searrow 0$.
But the entropy can nevertheless be defined by working with softened surface layer integrals
as first introduced in~\cite{linhyp}.
To this end, we choose a smooth function~$\eta \in C^\infty((t_0-\delta, t_0+\delta) \times M, \R)$
with~$0 \leq \eta \leq 1$ 
with the property that the function~$\theta(t,.) := \partial_t \eta(t,.)$ is non-negative.
We also write~$\eta(t,x)$ as~$\eta_t(x)$ and~$\theta(t,x)$ as~$\theta_t(x)$.
One can think of the function~$\eta_t$ as being identically equal to one in the distant past and
equal to zero in the distant future of the time~$t$. The support of~$\theta_t$ can be regarded as
a ``time strip'' localized near the time~$t$. For more details and further explanations
we refer to~\cite[Section~3.1]{linhyp} or~\cite[Section~6.2]{dirac}.

We define the {\em{softened nonlinear surface layer integral}} by modifying~\eqref{gammaeta} to
\begin{align*}
\gamma^{t,t'} \big( \eta \rho, \scrU (\eta \rho) \big) := &\int_M  d(\eta \rho)(x) \int_M d(\eta \rho)(y) \\
&\times \Big( \eta_t(x)\: \big(1-\eta_{t'}(y)\big) - \big(1-\eta_t(x)\big) \: \eta_{t'}(y)  \Big)
\L \big(x, \scrU y \scrU^{-1} \big) \:. 
\end{align*}
Then we can define the set~$\G^{t_0}(\Delta t)$ again by~\eqref{GtDt}.
Choosing~$\Delta t$ for which~$\G^{t_0}(\Delta t)$ is not a set of measure zero, we can define
the entropy~${\mathscr{S}}_{\tilde{\rho}, \eta \rho} \big(\Delta t, \tilde{\Omega} \big)$ again by~\eqref{SDO}.
This entropy is again non-negative. In general, the vacuum
entropy~${\mathscr{S}}_{\eta \rho, \eta \rho}(\Delta t, \Omega^{t_0})$ is not zero, but
an estimate similar to~\eqref{gttes} shows that this entropy becomes smaller if~$\Delta t$ is decreased.
}} \QEDrem
\end{Remark}

\subsection{The Case of a Static Vacuum} \label{secfinitestatic}
We now specialize the setting by restricting attention to vacuum spacetimes which
are static and time-periodic in the following sense.
\begin{Def} \label{deffinitestatic}
Let~$(U_t)_{t \in \R}$ be a one-parameter group of unitary transformations 
on the finite-dimensional Hilbert space~$\H$.
The causal fermion system~$(\H, \F, \rho)$ is {\bf{static and time-periodic}} with respect to~$(U_t)_{t \in \R}$
if it has the following properties:
\begin{itemize}[leftmargin=2em]
\item[\rm{(i)}] Spacetime $M:= \supp \rho \subset \F$ is a topological product,
\[ M = S^1 \times N \:. \]
We identify~$S^1$ with~$\R / (T_{\max} \Z)$ (where~$T_{\max}$ is again the life time
of one period of the universe) and write
a spacetime point~$x \in M$ as~$x=(t,\x)$ with~$t \in \R / (T_{\max} \Z)$ and~$\x \in N$.
\item[\rm{(ii)}] The one-parameter group~$(U_t)_{t \in \R}$ leaves
the measure~$\rho$ invariant, i.e.
\[ \rho\big( U_t \,\Omega\, U_t^{-1} \big) = \rho(\Omega) \qquad \text{for all
	$\rho$-measurable~$\Omega \subset \F$} \:. \]
Moreover,
\beq \label{txtrans}
U_{t'}\: (t,\x)\: U_{t'}^{-1} = (t+t',\x)\:.
\eeq
\end{itemize}
\end{Def} \noindent
Clearly, for static spacetimes, the measure~$\rho$ has again the representation~\eqref{rhodecomp},
where now~$d\mu_t=d\mu$ is time independent.

If the vacuum is static, the construction of the previous section simplifies because
one no longer needs to consider a time interval~$(t_0-\Delta t, t_0+\Delta t)$ 
and take the limit~$\Delta t \searrow 0$. Instead, one can work at fixed time, as we now explain.
We define the set
\[ 
\G^{t_0} := \big\{ \scrU \in \G \:\big|\: \gamma^{t_0, t_0} \big( \eta \rho, \scrU (\eta \rho) \big) = 0 \big\} 
\subset \G \:. \]
\begin{Lemma} The set~$\G^{t_0}$ is point-symmetric in the sense
\[ 
\scrU \in \G^{t_0} \quad \Longrightarrow \quad \scrU^{-1} \in \G^{t_0} \:. \]
\end{Lemma}
\Proof Using the symmetry properties of the nonlinear surface layer integral~\eqref{gammaeta}
and the unitary invariance of the Lagrangian,
\begin{align*}
\gamma^{t_0, t_0} \big( \eta \rho, \scrU (\eta \rho) \big) &= -\gamma^{t_0, t_0} \big( \scrU (\eta \rho), \eta \rho \big) \\
&= -\gamma^{t_0, t_0} \big( \scrU^{-1} \scrU (\eta \rho), \scrU^{-1} (\eta \rho) \big)
= -\gamma^{t_0, t_0} \big( \eta \rho, \scrU^{-1} (\eta \rho) \big) \:.
\end{align*}
Hence the left side vanishes if and only if the right side is zero. This gives the result.
\QED

Using~\eqref{txtrans}, the set~$\G^{t_0}(\Delta t)$ defined in~\eqref{GtDt} can be written simply as
\beq \label{Gt0D}
\G^{t_0}(\Delta t) = \big\{ \scrU U_\tau \:\big|\: \scrU \in \G^{t_0} ,\:|\tau| \leq \Delta t \big\} \:.
\eeq
Being defined by one real equation, one can hope that the set~$\G^{t_0}$ is a submanifold of~$\G$.
This is indeed the case under a technical assumption which we first define and explain afterward.
\begin{Def} \label{defttr}
The measure~$\eta \rho$ is {\bf{time translation regular}} at~$t_0$ if for all~$\scrU \in \G^{t_0}$,
\beq \label{regt0}
\int_{N_{t_0}} d\mu(x) \int_M \L \big( \scrU x \scrU^{-1}, y \big) \:d(\eta \rho)(y) > 0 \:.
\eeq
\end{Def} \noindent
This condition can be understood from the EL equations for minimizing measures, which
state that for all~$x \in \F$,
\[ 0 \leq \ell(x) = \int_M \L(x,y)\: d\rho(y) + \s \qquad \text{with} \qquad \s > 0 \:. \]
This inequality shows that, if in~\eqref{regt0} we replaced the measure~$d(\eta \rho)(y)$ 
in the inner integral by~$d\rho(y)$, then this integral and therefore also the whole expression
would be strictly positive. With the cutoff function~$\eta$ present, the inequality~\eqref{regt0}
is no longer obvious. But this inequality can be regarded as a condition for the choice of this
cutoff function. Again, this technical issue is a shortcoming of the finite-dimensional setting; it will
disappear in the infinite-dimensional setting of Section~\ref{secinfinite}.
\begin{Lemma} \label{lemmasub}
If the measure~$\eta \rho$ is time translation regular at~$t_0$,
then~$\G^{t_0}$ is a submanifold of~$\G$ of co-dimension one.
\end{Lemma}
\Proof We want to show that zero is a regular value of the function
\[ \phi \::\: \G \rightarrow \R \:, \qquad \phi(V) := \gamma^{t_0, t_0} \big( \eta \rho, V (\eta \rho) \big) \:. \]
To this end, it suffices to show that a specific directional derivative is non-zero, namely that
\[ 0 \neq \frac{d}{d\tau}\phi(V U_\tau) \big|_{\tau=0} =
\frac{d}{d\tau} \gamma^{t_0, t_0} \big( \eta \rho, V U_\tau (\eta \rho) \big) \Big|_{\tau=0} \:. \]
Using the time translation symmetry~\eqref{txtrans}, we have
\[ \gamma^{t_0, t_0} \big( \eta \rho, V U_\tau (\eta \rho) \big) =
\gamma^{t_0, t_0+\tau} \big( \eta \rho, V (\eta \rho) \big) \:, \]
making it possible to compute the $\tau$-derivative similar to~\eqref{gdiff},
\[ 
\frac{\partial}{\partial \tau} \gamma^{t_0, t_0+\tau} \big(\eta \rho, V (\eta \rho) \big) \Big|_{\tau=0}
= -\int_N d(\eta \mu)(x) \int_M \L \big(V x V^{-1}, y \big) \:d(\eta \rho)(y) \:. \]
Here we can replace~$\eta \mu$ by~$\mu$ because~$\eta$ is identically equal to one at time~$t_0$.
Therefore, the last integral is strictly negative by~\eqref{regt0}, concluding the proof.
\QED

From now on, we always assume that the measure~$\eta \rho$ is time translation regular at~$t_0$ (see Definition~\ref{defttr}). In the present finite-dimensional setting, the one-parameter unitary
group~$(U_\tau)_{\tau \in \R}$ can be clearly be written as
\beq \label{stonefinite}
U_\tau = e^{-i \tau H}
\eeq
with an infinitesimal generator~$H$, being a symmetric operator on the Hilbert space~$\H$.
This infinitesimal generator is a vector field on~$\G$, defined by the right-action
\[ H(f)(g) := \frac{d}{dt} f\big( g \,U_t \big) \big|_{t=0} \qquad \text{for~$g \in \G$, $f \in C^\infty(\G)$}\:. \]
The proof of Lemma~\ref{lemmasub} shows that this vector field is transversal to the submanifold~$\G^{t_0}$.
This makes it possible to introduce a canonical measure on~$\G^{t_0}$ by
\[ d\mu^{t_0}_\G = d\mu \lfloor H \:, \]
where the contraction can be written in local coordinates as
\[ d\mu^{t_0}_\G = \varepsilon_{i_1, \ldots, i_m}\: H^{i_1} \: dx^{i_2} \wedge \cdots \wedge dx^{i_{m}} \:, \]
where~$\varepsilon$ is the Levi-Civita tensor on~$\G$ and~$m= \dim \G$.

Now we can adapt the construction in Section~\ref{secfinitegen} by taking out~$\Delta t$
and replacing the measure~$d\mu_\G$ by~$d\mu_{\G^{t_0}}$.
\begin{Def} The pair $(\tilde{\Omega}, h)$ with a Borel subset~$\tilde{\Omega} \subset \tilde{M}$ and~$h \in \G$ is {\bf{admissible at fixed time}} if
\[ \fint_{\G^{t_0}} \gamma^{\tilde{\Omega}, t_0} \big(\tilde{\rho}, h \scrU (\eta \rho) \big)\:
d\mu^{t_0}_\G(\scrU) = 0 \:. \]
The set of admissible pairs is denoted by
\[ {\mathscr{A}}_{\tilde{\rho}, \eta \rho} \subset \mathfrak{P}(\tilde{M}) \times \G \:. \]
\end{Def}

\begin{Prp} \label{prpentropystatic}
If the vacuum spacetime is static and the measure~$\eta \rho$ is time translation regular at~$t_0$
(see Definition~\ref{defttr}), the entropy~${\mathscr{S}}_{\tilde{\rho}, \eta \rho}
\big(\tilde{\Omega} \big)$ in Definition~\ref{defentropyfinite} can be written as
\[ {\mathscr{S}}_{\tilde{\rho}, \eta \rho} \big(\tilde{\Omega} \big) = \inf_{h \in \G \,|\, (\tilde{\Omega}, h) \in
{\mathscr{A}}_{\tilde{\rho}, \eta \rho}}
{\mathscr{S}}_{\tilde{\rho}, \eta \rho}(h) \:, \]
where
\[ {\mathscr{S}}_{\tilde{\rho}, \eta \rho}(h) := \inf_{\tilde{\Omega}' \in \mathfrak{P}(\tilde{M}) \,|\, (\tilde{\Omega}', h) \in
{\mathscr{A}}_{\tilde{\rho}, \eta \rho}}
\log \fint_{ \G^{t_0}} e^{\beta \gamma^{\tilde{\Omega}', t_0} \big(\tilde{\rho}, h \scrU (\eta \rho) \big)}\:
d\mu^{t_0}_\G(\scrU) \:. \]
\end{Prp}
\Proof Using~\eqref{Gt0D}, we know that
\begin{align*}
\fint_{ \G^{t_0}(\Delta t)} e^{\beta \gamma^{\tilde{\Omega}', t_0} \big(\tilde{\rho}, h \scrU (\eta \rho) \big)}\:
d\mu_\G(\scrU) &= \frac{1}{2 \Delta t} \int_{-\Delta t}^{\Delta t} d\tau \fint_{ \G^{t_0}} d\mu^{t_0}_\G(\scrU)\:
e^{\beta \gamma^{\tilde{\Omega}', t_0} \big(\tilde{\rho}, h \scrU U_\tau (\eta \rho) \big)} \\
&= \frac{1}{2 \Delta t} \int_{-\Delta t}^{\Delta t} d\tau \fint_{ \G^{t_0}} d\mu^{t_0}_\G(\scrU)\:
e^{\beta \gamma^{\tilde{\Omega}', t_0+\Delta t} \big(\tilde{\rho}, h \scrU (\eta \rho) \big)}\:.
\end{align*}
Hence
\begin{align*}
&\fint_{ \G^{t_0}(\Delta t)} e^{\beta \gamma^{\tilde{\Omega}', t_0} \big(\tilde{\rho}, h \scrU (\eta \rho) \big)}\:
d\mu_\G(\scrU) - \fint_{ \G^{t_0}} e^{\beta \gamma^{\tilde{\Omega}', t_0} \big(\tilde{\rho}, h \scrU (\eta \rho) \big)}\:
d\mu^{t_0}_\G(\scrU) \\
&= \frac{1}{2 \Delta t} \int_{-\Delta t}^{\Delta t} d\tau \fint_{ \G^{t_0}} d\mu^{t_0}_\G(\scrU)\:
e^{\beta \gamma^{\tilde{\Omega}', t} \big(\tilde{\rho}, h \scrU (\eta \rho) \big)} \Big|_{t=t_0}^{t=t_0+\Delta t} \\
&\leq \Delta t \: \sup_{t \in [t_0-\Delta t, t_0+\Delta t]} \:\sup_{\scrU \in \G^{t_0}}
\bigg| \frac{\partial}{\partial t} e^{\beta \gamma^{\tilde{\Omega}', t} \big(\tilde{\rho}, h \scrU (\eta \rho) \big)} \bigg|
\leq C\: \Delta t\:,
\end{align*}
where in the last step we again used~\eqref{gdiff} together with the fact that~$M$ and~$\G^{t_0}$
are compact and that~$\L$ is continuous. This estimate shows that the limit~$\Delta t \searrow 0$
exists uniformly in~$(\tilde{\Omega}', h)$, giving the result.
\QED

\subsection{Existence of Optimal Configurations} \label{secexoptimal}
According to Proposition~\ref{prpentropystatic}, the entropy involves taking the infimum
over admissible pairs~$(h, \tilde{\Omega}')$ with~$h \in \G$ and $\tilde{\Omega}' \in {\mathfrak{P}}(\tilde{M})$. 
In this section we shall prove that the minimum is attained
for a specific choice of the past sets~${\mathfrak{P}}(\tilde{M})$.
To this end, we assume that~$\tilde{M}$ and the measure~$\tilde{\rho}$ admit a time splitting
\[ \tilde{M} = [0,T_{\max}] \times \tilde{N} \qquad \text{and} \qquad d\tilde{\rho} = r(t,\x)\: dt\: d\tilde{\mu}(\x) \:, \]
where~$r(t,\x)$ is a continuous function on~$\tilde{M}$ and~$\tilde{\mu}$ a measure on~$\tilde{N}$.
We define the past sets by
\beq \label{PMdef}
{\mathfrak{P}}(\tilde{M}) := \Big\{ (t,\x) \in \tilde{M} \:\big|\: t \leq T(\x) \text{ with }
T \in L^1\big(\tilde{N}, [0,T_{\max}] \big) \Big\} \:.
\eeq
For ease in notation, for a set~$\tilde{\Omega}'$ which is parametrized as in~\eqref{PMdef} by~$T$,
we replace the upper index~$\tilde{\Omega}'$ by the corresponding time function, i.e.\
\beq \label{Tnotation}
\gamma^{T, t_0}\big( \tilde{\rho}, h \scrU(\eta \rho) \big)
:= \gamma^{\tilde{\Omega}', t_0}\big( \tilde{\rho}, h \scrU(\eta \rho) \big) \:.
\eeq

\begin{Prp} \label{prpoptimal}
Given~$\tilde{\Omega}$ for which the entropy~${\mathscr{S}}_{\tilde{\rho}, \eta \rho}
\big(\tilde{\Omega} \big)$ is finite, there are~$h \in \G$ and~$\tilde{\Omega}' \in {\mathfrak{P}}(\tilde{M})$
such that both pairs~$(\tilde{\Omega}, h), (\tilde{\Omega}', h) \in {\mathscr{A}}_{\tilde{\rho}, \eta \rho}$
are admissible and
\beq \label{optimal}
{\mathscr{S}}_{\tilde{\rho}, \eta \rho} \big(\tilde{\Omega} \big) = \log \fint_{ \G^{t_0}} e^{\beta \gamma^{T, t_0} \big(\tilde{\rho}, h \scrU (\eta \rho) \big)}\: d\mu^{t_0}_\G(\scrU) \:.
\eeq
\end{Prp}
\Proof Considering a set~$\tilde{\Omega}'$ as in~\eqref{PMdef},
we can carry out the integrals over~$t$ and~$y$ to obtain
\[ \gamma^{\tilde{\Omega}^T, t}(\tilde{\rho}, \scrU \rho) = \int_{\tilde{N}} g(\x, T(x), \scrU)\: d\tilde{\mu}(\x) \]
with
\begin{align*}
g(\x, \tilde{t}, \scrU) := \bigg( \int_0^{\tilde{t}} \!\! r(t,\x)\: dt \int_{M \setminus \Omega} d\rho(y) - 
\int_{\tilde{t}}^{T_{\max}} \!\!\!\!\! r(t,\x)\: dt \int_{\Omega} d\rho(y) \bigg) 
\L \big( x,\scrU y \scrU^{-1} \big) \:.
\end{align*}
The function~$g$ is obviously continuous in all its arguments.

Let~$(h_n, T_n)$ with~$h_n \in \G$ and~$T_n \in L^1(\tilde{N}, [0,T_{\max}])$ be a minimizing sequence.
Since~$\G$ is compact, a subsequence of~$h_n$ converges in~$\G$.
Moreover, due to weak compactness of a closed ball of $L^1$-functions on a compact set, a subsequence
of~$T_n$ converges weakly
in~$L^1(\tilde{N}, [0,T_{\max}])$. Hence there is a subsequence with
\[ h_{n_\ell} \rightarrow h \qquad \text{and} \qquad T_{n_\ell} \rightharpoondown T \in L^1(\tilde{N}, [0,T_{\max}]) \:. \]
Then the sequence~$T_{n_\ell}$ converges pointwise almost everywhere (with respect to the measure~$\tilde{\mu}$).
Due to continuity of~$g$, also the function~$g(\x, T_{n_\ell}(x), h_{n_\ell} \scrU)$ converges almost everywhere.
Hence Lebesgue's dominated convergence theorem yields that
\[ \gamma^{\tilde{\Omega}^T_{n_\ell}, t}(\tilde{\rho}, h_{n_\ell} \scrU \rho) \rightarrow 
\gamma^{\tilde{\Omega}^T, t}(\tilde{\rho}, h \scrU \rho) \qquad
\text{for almost all~$\scrU \in \G$} \:. \]
Since~$\G$ is compact and all the functions are uniformly bounded,
we can again apply Lebesgue's dominated convergence theorem to conclude that
also the entropy converges. This concludes the proof.
\QED

\subsection{Characterization of Optimal Configurations}
Having proven that optimal configurations exist, we now derive the corresponding Euler-Lagrange equations
obtained by varying~$\tilde{\Omega}'$. To this end, we vary the time function~$T$ in~\eqref{PMdef}
by considering a variation~$(T_\tau)_{\tau \in [0, \delta)}$ for~$\delta>0$,
which depends smoothly on~$\tau$ and~$T_0(\x)=T(\x)$. Then,
again using the notation~\eqref{Tnotation},
\beq \label{tauderiv}
\frac{d}{d\tau} \gamma^{T_\tau, t_0}\big( \tilde{\rho}, h \scrU(\eta \rho) \big) \bigg|_{\tau=0} \!\!\!
= \int_{\tilde{N}_{t_0}} \bigg( \int_M \L \big( \scrU^{-1} h^{-1} x h \scrU, y \big) \:d(\eta \rho)(y) 
\bigg)\: \tilde{g}(\x)\: d\tilde{\mu}_{t_0}(x) \:,
\eeq
where we set~$d\tilde{\mu}_{t_0} = r(t_0,\x)\: d\tilde{\mu}(\x)$ and
\[ \tilde{g}(\x) := \frac{d}{d \tau} T_\tau(\x) \big|_{\tau=0} \:. \]
In order to treat the admissibility condition with Lagrange multipliers, we need to
impose a regularity condition similar to that in Definition~\ref{defttr}.
\begin{Def} \label{defttrtilde}
The measure~$\tilde{\rho}$ is {\bf{time translation regular}} for the pair~$(\tilde{\Omega}', h)$ if
for all~$\scrU \in \G^{t_0}$,
\beq \label{regt0tilde}
\int_{\tilde{N}_{t_0}} d\tilde{\mu}_{t_0}(x) \int_M \L \big( \scrU^{-1} h^{-1} x h \scrU, y \big) \:d(\eta \rho)(y) > 0 \:.
\eeq
\end{Def} \noindent

\begin{Prp} \label{lemmaELoptimal} Let~$h \in \G$ and~$\tilde{\Omega}' \subset M$ such
that~$(\tilde{\Omega}, h), (\tilde{\Omega}', h) \in {\mathscr{A}}_{\tilde{\rho}, \eta \rho}$ are admissible and
that the optimality property~\eqref{optimal} is satisfied. Moreover, assume that
the measure~$\tilde{\rho}$ is time translation regular for the pair~$(\tilde{\Omega}', h)$
(see Definition~\ref{defttrtilde}). Then there is a Lagrange parameter~$c \in \R$ such
that for all~$\x \in N$,
\[ \fint_{\G^{t_0}} \bigg( \int_M \L \big( \scrU^{-1} h^{-1} x h \scrU, y \big) \:d(\eta \rho)(y) \bigg)
\Big( e^{\beta \gamma^{T, t_0}\big( \tilde{\rho}, h \scrU(\eta \rho) \big)} - c \Big)\:
d\mu^{t_0}_{\G}(\scrU) = 0 \:. \]
\end{Prp}
\Proof We consider a smooth variation~$T_\tau$ for~$\tau \in [0, \delta)$ and~$\delta>0$.
Our first task is to show that the admissibility condition
\beq \label{admiss3}
\fint_{\G^{t_0}} \gamma^{T_\tau, t_0}\big( \tilde{\rho}, h \scrU(\eta \rho) \big)\: d\mu^{t_0}_{\G}(\scrU) = 0
\eeq
can be treated with a Lagrange multiplier term. To this end, we must verify that the constraint is
regular, meaning that there is a smooth variation with
\[ \frac{d}{d\tau} \fint_{\G^{t_0}} \gamma^{T_\tau, t_0}\big( \tilde{\rho}, h \scrU(\eta \rho) \big)\: d\mu^{t_0}_{\G}(\scrU)
\neq 0 \:. \]
We consider the variation
\[ T_\tau(\x) = T(\x) - \tau \:. \]
Then, similar to~\eqref{tauderiv},
\begin{align*}
&\frac{d}{d\tau} \fint_{\G^{t_0}} \gamma^{T-\tau, t_0}\big( \tilde{\rho}, h \scrU(\eta \rho) \big)\: d\mu^{t_0}_{\G}(\scrU) \\
&= -\fint_{\G^{t_0}} \bigg( \int_{\tilde{N}_{t_0}} \bigg( \int_M \L \big( \scrU^{-1} h^{-1} x h \scrU, y \big) \:d(\eta \rho)(y)
\bigg)\: d\tilde{\mu}_{t_0}(x) \bigg) \:d\mu^{t_0}_{\G}(\scrU) \:.
\end{align*}
Using the inequality~\eqref{regt0tilde}, we conclude that the resulting expression is strictly negative.
This shows that the constraint~\eqref{admiss3} is indeed regular.

Using the optimality and treating the constraint~\eqref{admiss3} with a Lagrange multiplier~$\beta c$, we obtain
\begin{align*}
0 &= \frac{d}{d\tau}
\fint_{ \G^{t_0}} \bigg( e^{\beta \gamma^{T_\tau, t_0} \big(\tilde{\rho}, h \scrU (\eta \rho) \big)}\:
- \beta c\: \gamma^{T_\tau, t_0} \big(\tilde{\rho}, h \scrU (\eta \rho) \big) \bigg)\:
d\mu^{t_0}_\G(\scrU) \\
&= \beta
\fint_{ \G^{t_0}} \Big( e^{\beta \gamma^{T_\tau, t_0} \big(\tilde{\rho}, h \scrU (\eta \rho) \big)} - c \Big)
\bigg( \frac{d}{d\tau} \gamma^{T_\tau, t_0}\big( \tilde{\rho}, h \scrU(\eta \rho) \big) \Big|_{\tau=0} \bigg)\:
d\mu^{t_0}_\G(\scrU) \:.
\end{align*}
Applying~\eqref{tauderiv} and using that the function~$\tilde{g}$ can be chosen arbitrarily, we obtain the result.
\QED

We finally remark that one could also consider variations of the group element~$h$.
This also gives rise to corresponding optimality conditions. Since the resulting computations
are rather involved, we shall not enter this analysis here.

\subsection{Uniqueness of Optimal Configurations}
Choosing~$\tilde{\rho}=\rho$ and~$\tilde{\Omega}=\Omega^{t_0}$,
the pair~$(\tilde{\Omega}, h=\e)$ is obviously an optimal configuration,
because the corresponding entropy is zero.
This raises the question whether the optimality determines~$\tilde{\Omega}$
uniquely, or whether there are other optimal configurations.
We again choose a smooth family of time functions
\beq \label{Ttaudef}
T_\tau : N \rightarrow [0, T_{\max}] \qquad \text{with} \qquad T_0 = t_0
\eeq
and introduce the past sets (this time in the vacuum spacetime)
\[ \Omega_\tau = \big\{ (t,\x) \in \R \times N \:\big|\: t \leq T_\tau(\x) \big\} \:. \]
Here we shall prove that, under certain assumptions, the
set~$\tilde{\Omega}$ is a strict local minimizer of the entropy (Theorem~\ref{thmunique}).
In order to motivate our assumptions, we return to the EL equations for the vacuum measure
as introduced in Section~\ref{secEL}. According to~\eqref{EL}, the function~$\ell : \F \rightarrow \R^+_0$
defined by~\eqref{elldef} is minimal on the support of~$\rho$. Typically, the function~$\ell$ is strictly positive
outside this support, implying that the dimension level set~$\ell^{-1}(0)$ is the same as that of spacetime.
In the examples of causal fermion systems constructed in Minkowski space
(see~\cite[Section~1.2]{cfs} or~\cite{oppio}), the dimension of spacetime is equal to four.
With this in mind, it is sensible to assume that the dimension of the above level set
is much smaller than the dimension of the Hilbert space~$\H$.
In our uniqueness theorem, we need a corresponding condition for the function~$\ell_\eta$
obtained by inserting the cutoff function~$\eta$,
\beq \label{elletadef}
\ell_\eta(x) := \int_M \L(x, y) \:d(\eta \rho)(y) - \s \:.
\eeq
As explained in the introduction and at the beginning of Section~\ref{secfinitegen},
this cutoff function is an artifact of the fact that spacetime is time-periodic.
If the parameter~$\delta$ is larger than the time range of the Lagrangian,
the functions~$\ell$ and~$\ell_\eta$ coincide in a neighborhood of~$N_{t_0}$.
This leads us to formulate the condition on the level set {\em{locally}} near~$x \in N_{t_0}$.
More precisely, we impose that there is a neighborhood~$U(x) \subset \F$ of~$x$ so that the Hausdorff
dimension of the level set intersected with~$U(x)$ is bounded from above by
\beq \label{levelset}
\dim \Big( \ell_\eta^{-1}\big( \ell_\eta(x) \big) \cap U  \Big) \leq 4n\, (f-7) - 2 \:.
\eeq
This condition must be satisfied for three points~$x_1, x_2, x_3 \in N_{t_0}$ which must be regular in the sense
that their spin spaces~$S_{x_i}:=x_i(\H)$ have maximal dimensions,
\beq \label{xireg}
\dim S_{x_i} = 2n \:.
\eeq
Moreover, the spin spaces must have pairwise trivial intersections,
\beq \label{Sintersect}
S_{x_i} \cap S_{x_j} = \{0\} \qquad \text{for all~$i \neq j$}\:.
\eeq
Finally, we need to assume that the parameter~$|\beta|$ in the formula for the entropy~\eqref{expint} is
sufficiently large (for a discussion of this point see Remark~\ref{rembeta}).
Here is the statement of our result.

\begin{Thm} \label{thmunique} Assume that the following conditions hold:
\bitem
\item[\rm{(i)}] There are three spacetime points~$x_1, x_2, x_3 \in N_{t_0}$ 
which are regular~\eqref{xireg} and whose
spin spaces have pairwise trivial intersections~\eqref{Sintersect}.
\item[\rm{(ii)}] The level sets through the points~$x_i$ of the function~$\ell_\eta$ defined by~\eqref{elletadef},
\[ \big\{ x \in \F \:\big|\:  \ell_\eta(x) = \ell_\eta(x_i) \big\} \]
have, locally near~$x_i$, the Hausdorff dimension at most~$4n\, (f-7) - 2$ (see~\eqref{levelset}).
\eitem
Then for every nontrivial variation~$T_\tau$~\eqref{Ttaudef}
and for sufficiently large~$|\beta|$, the second variation of the entropy is strictly positive.
\end{Thm}

We now enter the proof of this theorem, which will be completed at the end of this section.
In order to obtain an class of admissible variations, we shift the time functions~$T_\tau(\x)$ by a
time~$\Delta t$ by setting
\[  T^{\Delta t}_\tau(\x) := T_\tau(\x) - \Delta t(\tau) \]
with~$\Delta t$ a suitable function of~$\tau$. 
Then the admissibility condition reads
\beq \label{admiss2}
\fint_{\G^{t_0}} \gamma^{T^{\Delta t}_\tau, t_0}\big( \eta \rho, \eta \rho \big)\: d\mu^{t_0}_{\G}(\scrU)
= 0 \:.
\eeq
In order to prove that this condition can be satisfied for small~$\tau$, we first compute
the linearization in~$\Delta t$,
\begin{align*}
&\fint_{\G^{t_0}} \gamma^{T^{\Delta t}_\tau, t_0}\big( \eta \rho, \eta \rho \big)\: d\mu^{t_0}_{\G}(\scrU) 
- \fint_{\G^{t_0}} \gamma^{T_\tau, t_0}\big( \eta \rho, \eta \rho \big)\: d\mu^{t_0}_{\G}(\scrU) \notag \\
&= - \Delta t(\tau) \fint_{\G^{t_0}} d\mu^{t_0}_{\G}(\scrU)
\int_{N_{t_0}} \bigg( \int_M \L \big( x, y \big) \:d(\eta \rho)(y)
\bigg)\: d\mu(x) + \O\big( (\Delta t)^2 \big) \:,
\end{align*}
where the derivative was computed similar to~\eqref{gdiff}.
Using that the measure~$\eta \rho$ is translation regular~\eqref{regt0},
we conclude that the above linearization is strictly negative.
Therefore, the implicit function theorem yields the existence of a function~$\Delta t(\tau)$
with the property~\eqref{admiss2} for small~$\tau$.

Differentiating the corresponding surface layer integrals, we obtain
\beq \label{gdiff2}
\frac{d}{d\tau} \gamma^{\Omega^{\Delta t(\tau)}_\tau, \Omega}(\rho, \scrU \rho) \big|_{\tau=0}
= \int_N \big( g(\x) - c \big) \bigg( \int_M \L \big( \scrU x \scrU^{-1}, y \big) \:d(\eta \rho)(y) \bigg) \:d\mu(\x) \:,
\eeq
where
\[ g(\x) := \frac{d}{d \tau} T_\tau(\x) \big|_{\tau=0} \qquad \text{and} \qquad
c := \frac{d}{d \tau} (\Delta t)(\tau) \big|_{\tau=0} \:. \]

\begin{Lemma} Given a variation~$T_\tau$, the second variation of the entropy
vanishes for large~$\beta$ if and only if there are functions~$a : \G^{t_0} \rightarrow \R$
and~$b : N_{t_0} \rightarrow \R$ such that for all~$\scrU \in \G^{t_0}$ and $x \in N_{t_0}$,
\beq \label{abdef}
\int_M \L \big( \scrU x \scrU^{-1}, y \big) \:d(\eta \rho)(y) = a(\scrU)\, b(\x) \:.
\eeq
\end{Lemma}
\Proof Differentiating the constraint that the pair~$(\Omega^{\Delta t(\tau)}_{\tau}, \e)$ be admissible
gives the condition
\[\fint_{\G^{t_0}}  \frac{d}{d\tau} \gamma^{\Omega^{\Delta t(\tau)}_\tau, \Omega}(\rho, \scrU \rho) \Big|_{\tau=0}\:
d\mu^{t_0}(\scrU)  = 0 \:, \]
and using~\eqref{gdiff2} one finds that the parameter~$c$ is given explicitly by
\beq \label{cexplicit}
c = \frac{\displaystyle \int_N g(\x)\: h(\x)\: d\mu(\x)}{\displaystyle \int_N h(\x)\: d\mu(\x)} \:,
\eeq
where we introduced the functions
\[ h_\scrU(\x) := \int_M \L \big( \scrU x \scrU^{-1}, y \big) \:d(\eta \rho)(y) \qquad \text{and} \qquad
h(\x) := \fint_{\G^{t_0}} h_\scrU(\x) \: d\mu^{t_0}(\scrU) \:. \]

We next consider the exponential of the corresponding entropy
\[ {\mathscr{E}}\big( \Omega^{t(\tau)}_\tau \big) :=
\fint_{ \G^{t_0}} e^{\beta \gamma^{\Omega^{t(\tau)}_\tau, t_0} \big(\eta \rho, \scrU (\eta \rho) \big)}\:
d\mu^{t_0}_\G(\scrU) \:. \]
Its first and second variation are computed by
\begin{align*}
&\frac{d}{d\tau} {\mathscr{E}}\big( \Omega^{t(\tau)}_\tau \big) \Big|_{\tau=0} \notag \\
&= \beta \fint_{ \G^{t_0}} \bigg( \int_N \big( g(\x) - c \big) \: h_\scrU(\x) \:d\mu(\x) \bigg)\:
e^{\beta \gamma^{\Omega^{t(\tau)}_\tau, t_0} \big(\eta \rho, \scrU (\eta \rho) \big)}\: 
d\mu^{t_0}_\G(\scrU) \\
&\frac{d^2}{d\tau^2} {\mathscr{E}}\big( \Omega^{t(\tau)}_\tau \big) \Big|_{\tau=0} \notag \\
&= \beta^2 \fint_{ \G^{t_0}} \bigg( \int_N \big( g(\x) - c \big) \: h_\scrU(\x) \:d\mu(\x) \bigg)^2\,
e^{\beta \gamma^{\Omega^{t(\tau)}_\tau, t_0} \big(\eta \rho, \scrU (\eta \rho) \big)}\:
d\mu^{t_0}_\G(\scrU) + \O(\beta) \:. 
\end{align*}

Assume that the second variation vanishes to the highest order~$\sim \beta^2$. Then
\[ \int_N \big( g(\x) - c \big) \: h_\scrU(\x) \:d\mu(\x) = 0 \qquad \text{for all~$\scrU \in \G^{t_0}$}\:. \]
Using the explicit formula for~$c$ in~\eqref{cexplicit}, we conclude that
\[ \int_N g(\x) \,h_\scrU(\x) \:d\mu(\x) = \frac{\displaystyle \int_N g(\x)\:h(\x)\: d\mu(\x)}{\displaystyle \int_N h(\x)\: d\mu(\x)}\: \int_N h_\scrU(\x) \:d\mu(\x) \:. \]
Since the function~$g$ can be chosen arbitrarily, it follows that
\[ h_\scrU(\x) = h(\x)\: \frac{\displaystyle \int_N h_\scrU \:d\mu}{\displaystyle \int_N h\: d\mu} \:. \]
This means that the function~$h_\scrU(\x)$ can be written as the desired product~\eqref{abdef}.
\QED

\begin{Lemma} \label{lemmatrivimp}
Assume that there are three spacetime points~$x_1, x_2, x_3 \in N_{t_0}$ whose
spin spaces have pairwise trivial intersections~\eqref{Sintersect}.
Then for every symmetric operator~$A \in \Lin(\H)$ the following implication holds:
\begin{align}
\tr \Big( A\, \big(\1 - \pi_{x_i} \big)\, B \big(\1 - \pi_{x_i} \big) \Big) =0 \quad &\text{for all~$B \in \Lin(\H)$ symmetric
and~$i \in \{1,2,3\}$} \notag \\
&\quad \Longrightarrow \qquad A = 0 \:. \label{trivimp}
\end{align}
\end{Lemma}
\Proof We identify $\Lin(\H)$ with the tensor product~$\H \otimes \H$ via
\[ 
(u \otimes v) w := u\: \la v | w \ra_\H \:. \]
Let~$A$ be a symmetric operator satisfying the
condition on the left of~\eqref{trivimp}. Then for any~$i \in \{1,2,3\}$, the operator~$A$ lies in the subspace
\[ A \in E_i := \text{span} \big( \H \otimes S_{x_i}, S_{x_i} \otimes \H \big) \:. \]
Taking the intersection,
\[ A \in E := E_1 \cap E_2 \cap E_3 \:. \]
These intersections can be calculated inductively with the help of the computation rules
\begin{align*}
\text{span} \big( {\mathcal{U}},\: {\mathcal{V}} \big) \cap {\mathcal{W}} &=
\text{span} \big( {\mathcal{U}} \cap {\mathcal{W}},\: {\mathcal{V}} \cap {\mathcal{W}} \big) \\
\big(U_1 \otimes V_1 \big) \cap \big(U_2 \otimes V_2 \big) &=
\big(U_1 \cap U_2 \big) \otimes \big( V_1 \cap V_2 \big) \:.
\end{align*}
We find that~$E$ is spanned by spaces of the form~$U \otimes V$, where either~$U$ or~$V$ is
the intersection of at least two of the spin spaces~$S_{x_i}$. Using~\eqref{Sintersect}, we conclude
that either~$U$ or~$V$ is trivial. It follows that~$U \otimes V$ and thus also~$E$ are trivial.
Hence~$A$ is zero, concluding the proof.
\QED
We remark that the condition~\eqref{Sintersect} for three spacetime points could be weakened
to the condition that there must be $2k+1$ points~$\{x_1, \ldots, x_{2k+1} \}$ in~$N_{t_0}$ with~$k \geq 1$
such that for any~$k$ of these points~$\{y_1, \ldots, y_k\} \subset \{x_1, \ldots, x_{2k+1}\}$,
\[ \bigcap_{i=1}^k S_{y_i} = \{0\} \:. \]
We do not work out this refinement, because~\eqref{Sintersect} seems sufficient for the applications in mind.

For~$x \in M$ we define the compact subgroup
\[ \G_x := \big\{ \scrU \in \G \:|\: \scrU|_{S_x} = \1_{S_x} \big\} \subset \G \:. \]
\begin{Lemma} \label{lemmatransverse}
Assume that~\eqref{Sintersect} holds. Then there is~$i \in \{1,2,3\}$ such that~$\G^{t_0}$
and~$\G_{x_i}$ intersect transversely in~$\e$.
\end{Lemma}
\Proof According to Lemma~\ref{lemmasub}, $\G^{t_0}$ is a submanifold of~$\G$ of co-dimension one.
Identifying~$T_\e \G$ with the symmetric operators on~$\H$ and choosing the Hilbert-Schmidt scalar product,
we let~$\nu \neq 0$ be a normal to~$T_\e \G^{t_0}$, i.e.
\[ T_\e \G^{t_0} = \big\{\text{$A \in \Lin(\H)$ symmetric} \:\big|\: \tr ( \nu A ) = 0 \big\} \:. \]
On the other hand, the tangent space of the subgroups~$\G_{x_i}$ is given by
\beq \label{TeGform}
T_\e \G_{x_i} = \big\{(\1-\pi_{x_i})\, B\, (\1-\pi_{x_i})  \:\big|\: \text{$B \in \Lin(\H)$ symmetric} \big\} \:.
\eeq
If all these tangent spaces were subspaces of~$T_\e \G^{t_0}$, it would follow that
\[ \tr \big( \nu\, (\1-\pi_{x_i})\, B\, (\1-\pi_{x_i}) \big)=0 \]
for all~$i$ and~$B$. Lemma~\ref{lemmatrivimp} would imply that~$\nu$ vanishes, a contradiction.

We conclude that at least one of the tangent spaces~$T_\e\G_{x_i}$ is transverse to~$T_\e\G^{t_0}$.
\QED

\begin{Lemma} \label{lemmasubzero} Assume that~$x,y \in N_{t_0}$ are regular points in the sense that
\[ \dim S_x = \dim S_y=2n \:. \]
Moreover, assume that~$S_x \cap S_y = \{0\}$. Then the set
\[ \big\{ \scrU x \scrU^{-1} \text{ with } \scrU \in \G_y \big\} \subset \F \]
contains a submanifold~$\F_y(x)$ of~$\F$ with
\beq \label{dimlower}
\dim \F_y(x) \geq 4n\, (f-4) \qquad \text{and} \qquad x \in \F_y(x)
\eeq
(where~$f$ as in~\eqref{fdef}).
\end{Lemma}
\Proof We denote the unitary transformation of~$x$ with elements of~$\G_y$ by~$\Phi$,
\[ \Phi \::\: \G_y \rightarrow \F\:,\qquad \scrU \mapsto \scrU x \scrU^{-1} \:. \]
Its linearization takes the form
\[ D\Phi|_\e \::\: T_\e\G_y \rightarrow T_x\F\:,\qquad A \mapsto i [A, x] \:. \]

Let us estimate the rank of this linearization from below. Clearly, this rank is
greater or equal than the rank of the operator obtained by multiplying from the left
with~$(\1-\pi_x)$,
\beq \label{Asimp}
A \mapsto (\1-\pi_x) \:i [A,x] = i (\1-\pi_x) A x = (\1-\pi_x) \,(\1-\pi_y)\, A\, (\1-\pi_y) \,x
\eeq
(where in the last step we used the form of~$T_\e\G_y$ in~\eqref{TeGform}).
We next determine the rank of the operator products on the right and left:
Clearly, the rank of the operator on the left is bounded from below by
\[ \text{rank}\, (\1-\pi_x) \,(\1-\pi_y) \geq f - 2n - 2n = f-4n \:. \]
The operator~$(\1-\pi_y) x$, on the other hand, has rank~$2n$. Namely, otherwise there
would be a nonzero vector~$u \in S_x$ with
\[ 0 = (\1-\pi_y) \,\pi_x \, u = u - \pi_y u \:. \]
Then~$u$ would be an eigenvector of~$\pi_y$ of eigenvalue one, implying that~$y \in S_y$,
in contradiction to the assumption~$S_x \cap S_y = \{0\}$.

We conclude that the rank of the mapping~\eqref{Asimp} is at least as large as the
matrices with~$2n$ columns and~$f-4n$ rows. Since the matrix entries are complex
and the spaces spanning the columns and rows are orthogonal
(as being~$S_x$ and a subspace of~$S_x^\perp$, respectively), we obtain
\[ \text{rank}\, D\Phi|_\e \geq 4n\, (f-4n) \:. \]

In the final step we construct the desired submanifold as an immersion: We choose an~$4n\, (f-4n)$-dimensional
subspace~$I$ of~$T_\e \G_y$ such that the mapping~$D\Phi|_\e$ restricted to~$I$ is injective.
Next, we let~$\Omega \subset \G_y$ be the submanifold generated from a small
neighborhood~$B_\varepsilon(0) \subset I$ 
by applying the exponential map. Using the implicit function theorem, the mapping~$\Phi|_\Omega$
is an immersion. Hence~$\F_y(x):=\Phi(\Omega)$ is the desired submanifold.
\QED
We remark that the lower bound for the dimension in~\eqref{dimlower} may not be optimal,
but it is sufficient for our purposes.

\begin{Lemma} Assume that there are three spacetime points~$x_1, x_2, x_3 \in N_{t_0}$ whose
spin spaces have pairwise trivial intersections~\eqref{Sintersect}. Moreover, assume that the points are
regular~\eqref{xireg}.
Then there is a submanifold~$\F_{\min} \subset \F$ of dimension
\[ \dim \F_{\min} \geq 4n\, (f-7) - 1 \]
such that for suitable~$i \in \{1,2,3\}$, the function~$\ell_\eta$ defined in~\eqref{elletadef}
is constant of~$\F_{\min}$,
\[ \ell_\eta \big|_{\F_{\min}} \equiv \ell_\eta(x_i) \:. \]
\end{Lemma}
\Proof Let~$x_1, x_2, x_3 \in N_{t_0}$ such that~\eqref{Sintersect} holds.
According to Lemma~\ref{lemmatransverse},
there is~$i \in \{1,2,3\}$ such that the submanifolds~$\G^{t_0}$ and~$\G_{x_i}$ of~$\F$
intersect transversely at~$x_i$. Hence their intersection,
\[ \G^{t_0}_{x_i} := \G^{t_0} \cap \G_{x_i} \:, \]
locally near~$x_i$, is a submanifold of dimension
\beq \label{minusone}
\dim \big( \G^{t_0}_{x_i} \cap U(x_i) \big) = \dim \big( \G_{x_i} \cap U(x_i) - 1 \:.
\eeq
For all~$\scrU$ on this submanifold, the function~$a$ in~\eqref{abdef} is constant, because
\[ a(\scrU)\, b(x_i) = \ell_\eta \big(\scrU^{-1} x_i \scrU^{-1}\big) = \ell_\eta (x_i) = a(\e)\, b(x_i) \:. \]

Next we choose~$j \neq i$. According to Lemma~\ref{lemmasubzero} and using~\eqref{minusone}, the subset
\[ \big\{ \scrU^{-1} x_j \scrU \text{ with } \scrU \in \G^{t_0}_{x_i} \big\} \]
contains a submanifold of dimension~$4n (f-4)-1$. Since~$a$ is constant on~$\G^{t_0}_{x_i}$,
the function~$\ell_\eta$ is constant on this submanifold. This concludes the proof.
\QED

Theorem~\ref{thmunique} follows immediately from the last lemma.

\section{The Infinite-Dimensional Setting with Static Vacuum} \label{secinfinite}
We now turn attention to the infinite-dimensional case.
This has the advantage that the vacuum spacetime~$M$ can be chosen to have infinite lifetime,
so that the cutoff function~$\eta$ in~\eqref{gammaeta} is no longer needed.
However, the complication arises that the unitary group~$\U(\H)$ is infinite-dimensional,
making it necessary to exhaust this group by finite-dimensional subgroups.

Let~$(\H, \F, \rho)$ be a causal fermion system describing the vacuum.
We assume that the system is is static.
In contrast to the finite-dimensional setting (see Definition~\ref{deffinitestatic}), we
can now consider a spacetime of infinite lifetime. Moreover, we need to
impose that the one-parameter family of time translations be strongly continuous.
\begin{Def} \label{defstatic}
Let~$(U_t)_{t \in \R}$ be a strongly continuous one-parameter group of unitary transformations 
on the Hilbert space~$\H$ (i.e.\ $s$-$\lim_{t' \rightarrow t} U_{t'}= U_t$ and~$U_t U_{t'} = U_{t+t'}$).
The causal fermion system~$(\H, \F, \rho)$ is {\bf{static with respect to~$(U_t)_{t \in \R}$}}
if it has the following properties:
\begin{itemize}[leftmargin=2em]
\item[\rm{(i)}] Spacetime $M:= \supp \rho \subset \F$ is a topological product,
\[ M = \R \times N \:. \]
We write a spacetime point~$x \in M$ as~$x=(t,\x)$ with~$t \in \R$ and~$\x \in N$.
\item[\rm{(ii)}] The one-parameter group~$(U_t)_{t \in \R}$ leaves
the measure~$\rho$ invariant, i.e.
\[ \rho\big( U_t \,\Omega\, U_t^{-1} \big) = \rho(\Omega) \qquad \text{for all
	$\rho$-measurable~$\Omega \subset \F$} \:. \]
Moreover,
\[ U_{t'}\: (t,\x)\: U_{t'}^{-1} = (t+t',\x)\:. \]
\end{itemize}
\end{Def} \noindent
Using Stone's theorem (see for example~\cite[Theorem~VIII.8]{reed+simon}),
we can write the group action similar to~\eqref{stonefinite} as an exponential
\[ U_t = e^{-i t H} \:, \]
where the infinitesimal generator~$H$ is a selfadjoint operator on~$\H$
with dense domain denoted by~$\D(H)$.

Next, we let~$(\tilde{\H}, \tilde{\F}, \tilde{\rho})$ be a causal fermion system describing the interacting system.
We again assume that both measures~$\rho$ and~$\tilde{\rho}$ are minimizers of the causal action.
After identifying the two Hilbert spaces via a unitary mapping~$V : \H \rightarrow \tilde{\H}$,
we can work exclusively in the Hilbert space~$\H$. However, we must keep in mind that this
identification is not canonical, leaving us again with the freedom to transform~$V$ according to~\eqref{Vtrans}.

For any~$t \in \R$, we let~$\Omega^t$ be the past of~$t$,
\[ 
\Omega^t := \{ x \in M \:|\: \scrt(x) \leq t \} \:. \]
We choose~$\Omega:=\Omega^{t_0}$ as the past of some fixed time~$t_0$.
Given a Borel set~$\tilde{\Omega} \subset \tilde{M}$, the {\em{nonlinear surface layer integral}}
is defined formally by
\beq \label{osinlformal}
\begin{split}
\gamma^{\tilde{\Omega}, \Omega}(\tilde{\rho}, \scrU \rho) &= \int_{\tilde{\Omega}} d\tilde{\rho}(x) \int_{M \setminus \Omega} d\rho(y)\:
\L \big(x,\scrU y \scrU^{-1} \big) \\
&\quad\, - \int_{\tilde{M} \setminus \tilde{\Omega}} d\tilde{\rho}(x) \int_{\Omega} d\rho(y)\:
\L \big(x,\scrU y \scrU^{-1} \big)\:.
\end{split}
\eeq
Clearly, we need to make sure that these integrals converge.
For our purposes, it is most convenient to work with the following rather weak notion of convergence.

\begin{Def} \label{defic}
The surface layer integral~$\gamma^{\tilde{\Omega}, \Omega}(\tilde{\rho}, \scrU \rho)$ is
{\bf{conditionally convergent}} if for any~$\y \in N$, the following integrals are finite,
\[ \int_{\tilde{\Omega}} d\tilde{\rho}(x) \int_{t_0}^\infty dt\:
\L \big(x,\scrU \,(t,\y)\, \scrU^{-1} \big),\; \int_{\tilde{M} \setminus \tilde{\Omega}} d\tilde{\rho}(x) \int_{-\infty}^{t_0} dt\:
\L \big(x,\scrU \,(t,\y)\, \scrU^{-1} \big) < \infty \:, \]
and if
\begin{align*}
\int_N d\mu(\y) \:\bigg| &\int_{\tilde{\Omega}} d\tilde{\rho}(x) \int_{t_0}^\infty dt\:
\L \big(x,\scrU \,(t,\y)\, \scrU^{-1} \big) \\
&- \int_{\tilde{M} \setminus \tilde{\Omega}} d\tilde{\rho}(x) \int_{-\infty}^{t_0} dt\:
\L \big(x,\scrU \,(t,\y)\, \scrU^{-1} \big) \bigg| < \infty \:.
\end{align*}
\end{Def} \noindent
Under these assumptions, we can define the nonlinear surface layer integral by
\begin{align*} 
\gamma^{\tilde{\Omega}, \Omega}(\tilde{\rho}, \scrU \rho) :=
\int_N d\mu(\y) \:\bigg( &\int_{\tilde{\Omega}} d\tilde{\rho}(x) \int_{t_0}^\infty dt\:
\L \big(x,\scrU \,(t,\y)\, \scrU^{-1} \big) \\
&- \int_{\tilde{M} \setminus \tilde{\Omega}} d\tilde{\rho}(x) \int_{-\infty}^{t_0} dt\:
\L \big(x,\scrU \,(t,\y)\, \scrU^{-1} \big) \bigg) \:.
\end{align*}

We choose a finite-dimensional subspace of the domain of~$H$,
\[ \H^\fermi \subset \D(H) \subset \H \:, \]
and let~$\G^\fermi := \U(\H^\fermi)$ be the corresponding unitary group.
\begin{Def} The group~$\G^\fermi$ is {\bf{admissible}} if the subset
\beq \label{Gt0definf}
\G^{\fermi, t_0} := \big\{ \scrU \in \G^\fermi \:\big|\: \gamma^{t_0, t_0} \big( \rho, \scrU \rho \big) 
\text{ is conditionally convergent and vanishes} \big\}
\eeq
is a co-dimension one submanifold of~$\G^\fermi$.
\end{Def}
The symmetric operator~$H_{\G^\fermi} := \pi_{\H^\fermi} H \pi_{\H^\fermi}$
can be viewed as a vector field on~$\G^\fermi$. This makes it possible to define a canonical measure
on~$\G^{\fermi, t_0}$ by
\[ d\mu^{t_0}_{\G^\fermi} := d\mu \lfloor H_{\G^\fermi} \]
(we do not need that~$H_{\G^\fermi}$ is transverse on~$\G^{\fermi, t_0}$).
Similar as in the finite-dimensional setting~\eqref{Pfinite}, we choose the {\em{past sets}}~${\mathfrak{P}}(\tilde{M})$
as a subset of the Borel sets of~$\tilde{M}$,
\[ 
{\mathfrak{P}}(\tilde{M}) \subset {\mathfrak{B}}(\tilde{M}) \:. \]

\begin{Def} The pair $(\tilde{\Omega}, h)$ with a Borel subset~$\tilde{\Omega} \subset \tilde{M}$ and~$h \in \G^\fermi$ is {\bf{admissible}} if the surface layer integral~$\gamma^{\tilde{\Omega}, t_0} (\tilde{\rho}, h \scrU \rho)$
is conditionally convergent for all~$\scrU \in \G^{\fermi, t_0}$ and if
\beq \label{adminfinite}
\fint_{\G^{\fermi, t_0}} \gamma^{\tilde{\Omega}, t_0} \big(\tilde{\rho}, h \scrU \rho \big)\:
d\mu^{t_0}_{\G^\fermi}(\scrU) =0 \:.
\eeq
The set of admissible pairs is denoted by
\[ {\mathscr{A}}_{\tilde{\rho}, \rho} \subset \mathfrak{P}(\tilde{M}) \times \G^\fermi \:. \]
\end{Def}

\begin{Def} \label{defSinf} The entropy~${\mathscr{S}}_{\tilde{\rho}, \rho}$ is defined by
\[ {\mathscr{S}}_{\tilde{\rho}, \rho} \big(\tilde{\Omega} \big) = 
\liminf_{\H^\fermi \nearrow \H}
\inf_{h \in \G^\fermi \,|\, (\tilde{\Omega}, h) \in
{\mathscr{A}}_{\tilde{\rho}, \rho}}
{\mathscr{S}}_{\tilde{\rho}, \rho}(h) \:, \]
where
\[ {\mathscr{S}}_{\tilde{\rho}, \rho}(h) := \inf_{\tilde{\Omega}' \in \mathfrak{P}(\tilde{M}) \,|\, (\tilde{\Omega}', h) \in
{\mathscr{A}}_{\tilde{\rho}, \rho}}
\log \fint_{ \G^{\fermi, t_0}} e^{\beta \gamma^{\tilde{\Omega}', t_0} \big(\tilde{\rho}, h \scrU \rho \big)}\:
d\mu^{t_0}_{\G^\fermi}(\scrU) \:. \]
\end{Def}

\begin{Thm} The entropy is non-negative, i.e.\ for all Borel subsets~$\tilde{\Omega} \subset \tilde{M}$,
\[ {\mathscr{S}}_{\tilde{\rho}, \rho}(\tilde{\Omega}) \geq 0 \:. \]
Moreover, the entropy vanishes in the vacuum at time~$t_0$, i.e.
\beq \label{Svacinf}
{\mathscr{S}}_{\rho, \rho}(\Omega^{t_0}) = 0 \:.
\eeq
\end{Thm}
\Proof The non-negativity follows immediately from~\eqref{adminfinite} and Jensen's inequality.
In order to prove~\eqref{Svacinf} we choose~$\tilde{\rho}=\rho$, $h=\e$ and~$\tilde{\Omega}=\tilde{\Omega}'=\Omega^{t_0}$. Then by definition~\eqref{Gt0definf}, it follows immediately that
\[ \fint_{\G^{\fermi, t_0}} \gamma^{t_0, t_0} \big(\rho, \scrU \rho \big)\: d\mu^{t_0}_{\G^\fermi}(\scrU) =0 \:, \]
showing that the pair~$(\Omega^{t_0}, \e)$ is admissible. Hence
\[ {\mathscr{S}}_{\rho, \rho}(\Omega^{t_0}) \leq 
\log \fint_{ \G^{\fermi, t_0}} e^{\beta \gamma^{t_0, t_0} \big(\rho,\scrU \rho \big)}\:
d\mu^{t_0}_{\G^\fermi}(\scrU) = 0 \:, \]
concluding the proof.
\QED

\section{A Corresponding Localized Entropy and Entanglement Entropy} \label{secloc}
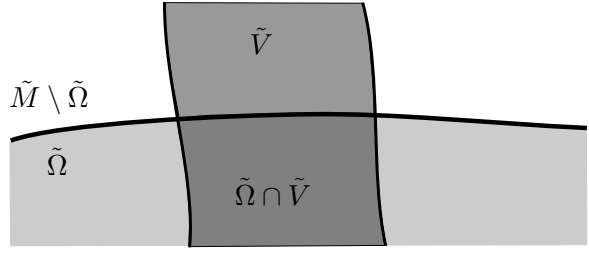
\begin{figure}[tb]
\psscalebox{1.0 1.0} 
{
\begin{pspicture}(0,26.420364)(7.692473,29.705011)
\definecolor{colour0}{rgb}{0.8,0.8,0.8}
\definecolor{colour1}{rgb}{0.6,0.6,0.6}
\definecolor{colour3}{rgb}{0.5019608,0.5019608,0.5019608}
\pspolygon[linecolor=colour0, linewidth=0.002, fillstyle=solid,fillcolor=colour0](0.01100586,27.8297)(0.017959623,26.452324)(7.681006,26.465631)(7.6740522,28.002705)(6.582311,28.055937)(5.2889113,28.149092)(4.426645,28.202324)(3.4253032,28.202324)(2.2848861,28.135784)(1.4365271,28.069244)(0.504723,27.949472)
\pspolygon[linecolor=black, linewidth=0.002, fillstyle=solid,fillcolor=colour1](2.0692356,29.687325)(4.70402,29.679234)(4.761939,29.458246)(4.791628,29.144083)(4.827932,28.71211)(4.854391,28.39595)(4.851006,28.203955)(4.258745,28.2155)(3.3673575,28.19432)(2.6711853,28.188505)(2.2616878,28.152325)(2.157154,28.702475)(2.0858502,29.186531)(2.076006,29.517153)
\pspolygon[linecolor=black, linewidth=0.002, fillstyle=solid,fillcolor=colour3](2.2460058,28.12765)(3.2003558,28.188513)(4.1024127,28.21556)(4.8591266,28.212324)(4.8737364,27.870676)(4.886775,27.38378)(4.9210296,27.070946)(4.971786,26.673723)(5.011006,26.457325)(2.3909822,26.44085)(2.400592,26.808971)(2.3659823,27.28058)(2.3052256,27.761002)(2.2525425,27.985638)
\psbezier[linecolor=black, linewidth=0.04](4.701006,29.687325)(4.9610057,28.686077)(4.7960057,27.404078)(5.006006,26.45232421875)
\psbezier[linecolor=black, linewidth=0.04](2.066006,29.694824)(2.0559301,28.619019)(2.4891877,27.353392)(2.398506,26.43232421875)
\psbezier[linecolor=black, linewidth=0.06](0.021373715,27.852324)(0.6836917,28.084381)(2.8376231,28.18811)(4.0002766,28.20234561418121)(5.16293,28.21658)(6.2635965,28.088652)(7.681006,28.01923)
\rput[bl](0,28.2){$\tilde{M} \setminus \tilde{\Omega}$}
\rput[bl](0.5,27.4){$\tilde{\Omega}$}
\rput[bl](3.2,29){$\tilde{V}$}
\rput[bl](3,27){$\tilde{\Omega} \cap \tilde{V}$}
\end{pspicture}
}
\caption{Localizing the entropy.}
\label{figcylinder}
\end{figure}

The above notion of entropy can be ``localized'' such as to obtain the entropy of a
subset of space. Moreover, there is a corresponding notion of entanglement entropy.
In order to describe the spatial region, we choose a subset~$\tilde{V} \subset \tilde{M}$ which can be
thought of as a ``cylinder'' in spacetime (see Figure~\ref{figcylinder}).
We ``localize''
the nonlinear surface layer integral~\eqref{osinlformal} by restricting the $\tilde{\rho}$-integrals to~$\tilde{V}$,
\beq \label{osinlocal}
\begin{split}
\gamma^{\tilde{\Omega}, \Omega}_{\tilde{V}}
(\tilde{\rho}, \scrU \rho) &= \int_{\tilde{\Omega} \cap \tilde{V}} d\tilde{\rho}(x) \int_{M \setminus \Omega} d\rho(y)\:
\L \big(x,\scrU y \scrU^{-1} \big) \\
&\quad\, - \int_{(\tilde{M} \setminus \tilde{\Omega}) \cap \tilde{V}} d\tilde{\rho}(x) \int_{\Omega} d\rho(y)\:
\L \big(x,\scrU y \scrU^{-1} \big)\:.
\end{split}
\eeq
Obviously, this expression is additive in~$\tilde{V}$ in the sense that for two
disjoint subsets~$\tilde{V}, \tilde{V}' \subset \tilde{M}$,
\[ \gamma^{\tilde{\Omega}, \Omega}_{\tilde{V} \cup \tilde{V}'}(\tilde{\rho}, \scrU \rho)
= \gamma^{\tilde{\Omega}, \Omega}_{\tilde{V}} (\tilde{\rho}, \scrU \rho)
+ \gamma^{\tilde{\Omega}, \Omega}_{\tilde{V}'} (\tilde{\rho}, \scrU \rho) \:. \]
Improper convergence of these integrals can be defined in analogy to Definition~\ref{defic}.
We denote the admissible pairs for which also the localized surface layer integrals
converge conditionally by~${\mathscr{A}}^V_{\tilde{\rho}, \rho}$.
The {\em{localized entropy}} of~$\tilde{V}$ is defined by
\[ {\mathscr{S}}_{\tilde{\rho}, \rho} \big(\tilde{\Omega} , \tilde{V} \big) = 
\liminf_{\H^\fermi \nearrow \H}
\inf_{h \in \G^\fermi \,|\, (\tilde{\Omega}, h) \in
{\mathscr{A}}^V_{\tilde{\rho}, \rho}}
{\mathscr{S}}_{\tilde{\rho}, \rho}(h, \tilde{V}) \:, \]
where
\[ {\mathscr{S}}_{\tilde{\rho}, \rho} \big(h, \tilde{V} \big) := \inf_{\tilde{\Omega}' \in \mathfrak{P}(\tilde{M}) \,|\, (\tilde{\Omega}', h) \in
{\mathscr{A}}^V_{\tilde{\rho}, \rho}}
\log \fint_{ \G^{\fermi, t_0}} e^{\beta \gamma^{\tilde{\Omega}', t_0}_{\tilde{V}} \big(\tilde{\rho}, h \scrU \rho \big)}\:
d\mu^{t_0}_{\G^\fermi}(\scrU) \:. \]
The {\em{entanglement entropy}} of~$\tilde{V}$ is defined by
\[ {\mathscr{E}}_{\tilde{\rho}, \rho} \big(\tilde{\Omega} , \tilde{V} \big) :=
{\mathscr{S}}_{\tilde{\rho}, \rho} \big(\tilde{\Omega}\big)
- {\mathscr{S}}_{\tilde{\rho}, \rho} \big(\tilde{\Omega} , \tilde{V} \big) 
- {\mathscr{S}}_{\tilde{\rho}, \rho} \big(\tilde{\Omega} , \tilde{M} \setminus \tilde{V} \big) \:. \]
At present, it is not known whether the localized entropy has subadditivity properties.
Likewise, it is unknown if the entanglement entropy is always non-negative.

\section{Comparison and Outlook} \label{secoutlook}

\begin{Remark} \label{rembeta}
{\bf{(Significance of the parameter~$\beta$)}} {\em{
Our entropy involves a real parameter~$\beta$ which appears in the exponential of the
defining equation~\eqref{expint}. In order to understand the significance of this parameter,
we first determine its length dimension. Having fixed the local trace~\eqref{fixedtrace}, the Lagrangian
is dimensionless. Consequently, being a double integral over spacetime, the nonlinear
surface layer integral~\eqref{OSInonlin} has length dimension eight.
It involves fluctuations on different length scales (as described by so-called holographic components; for details see~\cite{mix}).
Then the scaling behavior tells us that the fluctuations of size~$d$ scale like~$d^{-8}$.
Since the argument of the exponential must be dimensionless, we conclude that the parameter~$\beta$
has length dimension minus eight. Hence the exponential takes into account mainly the fluctuations
on the scale
\[ d \gtrsim |\beta|^{-\frac{1}{8}} \:. \]
Therefore, the parameter~$\beta^{-\frac{1}{8}}$ can
be regarded as the length scale of an ultraviolet cutoff for the fluctuations to be taken into
account by the entropy. Considering a large value of~$|\beta|$ (as done in
the uniqueness result of Theorem~\ref{thmunique}) corresponds to taking into account
small-scale fluctuations.
}} \QEDrem
\end{Remark}

\begin{Remark} \label{remcompare}
{\bf{(Connection to other notions of entropy)}} {\em{
The goal of this paper was to define a general notion of entropy for causal fermion systems
and to analyze a few basic properties. The connection to other notions of entropy
(in particular to the von Neumann entropy and the corresponding entanglement entropy)
is largely unknown and remains an interesting topic for future research.

Here we conclude with a short comparison to a notion of von Neumann entropy
for causal fermion systems:
Representing the quantum state~$\omega^t$ constructed in~\cite{fockfermionic} gives rise to a density
operator~$\sigma^t$ on a Fock space~$\Fock$ (see~\cite[Section~4.5]{fockfermionic}).
This makes it possible to define the corresponding von Neumann entropy by
\[ {\mathscr{S}}^t_{\text{vN}} := -\tr \big( \sigma^t\, \log \sigma^t \big) \:. \]
Although formulated with similar notions, the connection between this entropy
and the entropy~${\mathscr{S}}(\tilde{\Omega})$ defined above is unclear.
The main difficulty in analyzing this connection is that~$\sigma^t$, and consequently also the
von Neumann entropy, depend on the choice of representation, whereas~${\mathscr{S}}(\tilde{\Omega})$
is independent of Fock representations. This suggests that the two entropies do not coincide.
It seems that the entropy~${\mathscr{S}}(\tilde{\Omega})$ is more general and more universal.
}} \QEDrem
\end{Remark}

\Thanks{{{\em{Acknowledgments:}} I would like to thank Jos{\'e} Isidro, Niky Kamran and Magdalena Lottner
for helpful discussions. I am grateful to the referees for valuable feedback and suggestions.

\providecommand{\bysame}{\leavevmode\hbox to3em{\hrulefill}\thinspace}
\providecommand{\MR}{\relax\ifhmode\unskip\space\fi MR }
\providecommand{\MRhref}[2]{%
  \href{http://www.ams.org/mathscinet-getitem?mr=#1}{#2}
}
\providecommand{\href}[2]{#2}

\end{document}